\documentclass[sigconf]{acmart}

\AtBeginDocument{%
  \providecommand\BibTeX{{%
    \normalfont B\kern-0.5em{\scshape i\kern-0.25em b}\kern-0.8em\TeX}}}

\copyrightyear{2024} 
\acmYear{2024} 
\setcopyright{rightsretained} 
\acmConference[CHI '24]{Proceedings of the CHI Conference on Human Factors in Computing Systems}{May 11--16, 2024}{Honolulu, HI, USA}
\acmBooktitle{Proceedings of the CHI Conference on Human Factors in Computing Systems (CHI '24), May 11--16, 2024, Honolulu, HI, USA}
\acmDOI{10.1145/3613904.3642579}
\acmISBN{979-8-4007-0330-0/24/05}

\usepackage{bm} 
\usepackage{float}
\usepackage{algorithm2e} 
\RestyleAlgo{ruled}
\SetKwComment{Comment}{/* }{ */}
\SetKwInOut{Require}{Require}
\SetKwInOut{Input}{Input}

\newcommand\ind{\protect\mathpalette{\protect\independenT}{\perp}}

\def\independenT#1#2{\mathrel{\rlap{$#1#2$}\mkern2mu{#1#2}}}
\definecolor{highlightcolor}{rgb}{1, 1, 0.6} 

\newcommand{\sout}[1] {}

\newcommand{\st}[1] {}

\usepackage{listings}
\definecolor{codegreen}{rgb}{0,0.6,0}
\definecolor{codegray}{rgb}{0.5,0.5,0.5}
\definecolor{codepurple}{rgb}{0.58,0,0.82}
\definecolor{backcolour}{rgb}{0.95,0.95,0.92}
\definecolor{backcolour_prompts}{rgb}{0.92,0.91,0.62}

\lstdefinestyle{mystyle}{
    backgroundcolor=\color{backcolour},   
    commentstyle=\color{codegreen},
    keywordstyle=\color{magenta},
    numberstyle=\tiny\color{codegray},
    stringstyle=\color{codepurple},
    basicstyle=\ttfamily\footnotesize,
    breakatwhitespace=false,         
    breaklines=true,                 
    captionpos=b,                    
    keepspaces=true,                 
    numbersep=5pt,                  
    showspaces=false,                
    showstringspaces=false,
    showtabs=false,                  
    tabsize=2
}

\lstdefinestyle{prompts}{
    backgroundcolor=\color{backcolour_prompts},   
    commentstyle=\color{codegreen},
    keywordstyle=\color{magenta},
    numberstyle=\tiny\color{codegray},
    stringstyle=\color{codepurple},
    basicstyle=\ttfamily\footnotesize,
    breakatwhitespace=false,         
    breaklines=true,                 
    captionpos=b,                    
    keepspaces=true,                 
    numbersep=5pt,                  
    showspaces=false,                
    showstringspaces=false,
    showtabs=false,                  
    tabsize=2
}

\begin{document}

\title[LLMR: Real-time Prompting of Interactive Worlds using Large Language Models]{LLMR: Real-time Prompting of Interactive Worlds  \protect\\using Large Language Models}

\author{Fernanda De La Torre}
\authornote{Authors contributed equally to this research and were affiliated with Microsoft.}
\email{dlatorre@mit.edu}
\affiliation{%
  \institution{Massachusetts Institute of Technology}
  \country{USA}
}

\author{Cathy Mengying Fang}
\authornotemark[1]
\email{catfang@media.mit.edu}
\affiliation{%
  \institution{MIT Media Lab}
  \country{USA}
}

\author{Han Huang}
\authornotemark[1]
\email{huangh14@rpi.edu}
\affiliation{%
  \institution{Rensselaer Polytechnic Institute}
  \country{USA}
}

\author{Andrzej Banburski-Fahey}
\email{abanburski@microsoft.com}
\affiliation{%
  \institution{Microsoft}
  \country{USA}
}

\author{Judith Amores Fernandez}
\email{judithamores@microsoft.com}
\affiliation{%
  \institution{Microsoft}
  \country{USA}
}

\author{Jaron Lanier}
\email{jalani@microsoft.com}
\affiliation{%
  \institution{Microsoft}
  \country{USA}
}

\renewcommand{\shortauthors}{De La Torre, Fang, and Huang et al.}
\newcommand{\framework}{LLMR}

\begin{abstract}

We present Large Language Model for Mixed Reality (LLMR), a framework for the real-time creation and modification of interactive Mixed Reality experiences using LLMs. LLMR leverages novel strategies to tackle difficult cases where ideal training data is scarce, or where the design goal requires the synthesis of internal dynamics, intuitive analysis, or advanced interactivity. Our framework relies on text interaction and the Unity game engine. By incorporating techniques for scene understanding, task planning, self-debugging, and memory management, LLMR outperforms the standard GPT-4 by 4x in average error rate. We demonstrate LLMR's cross-platform interoperability with several example worlds, and evaluate it on a variety of creation and modification tasks to show that it can produce and edit diverse objects, tools, and scenes. Finally, we conducted a usability study (N=11) with a diverse set that revealed participants had positive experiences with the system and would use it again.

\end{abstract}

\ccsdesc[500]{Computing methodologies~Spatial and physical reasoning}
\ccsdesc[500]{Computing methodologies~Multi-agent systems}

\keywords{large language model, mixed reality, spatial reasoning, artificial intelligence}


\begin{teaserfigure}
      \centering
      \includegraphics[width=.9\textwidth]{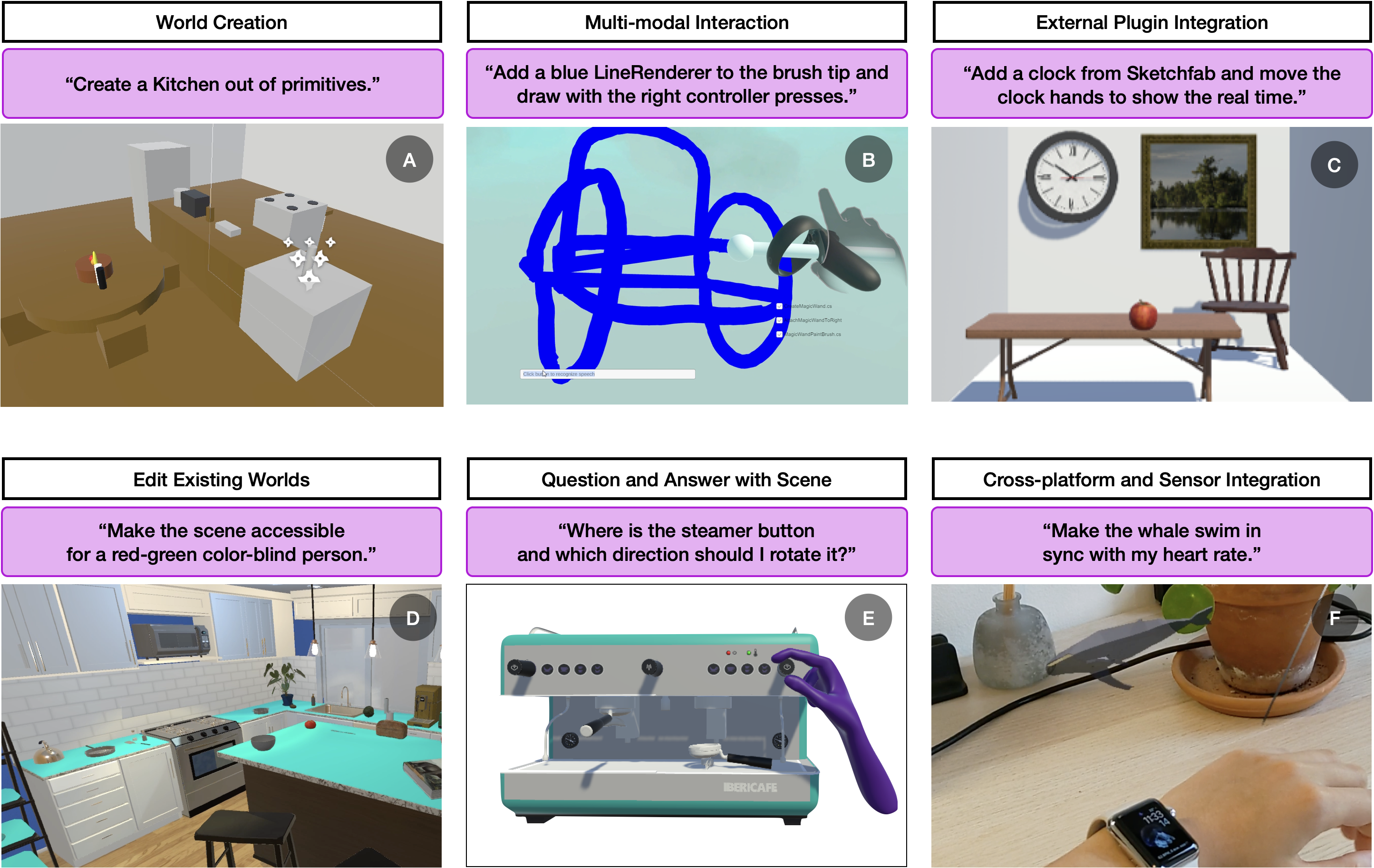}
      \caption{Examples of diverse use cases and functionalities enabled by the Large Language Model for Mixed Reality (\textit{\framework}) framework. A: Creation of a detailed kitchen scene from scratch using Unity primitives. B: Prompting and drawing objects into existence via multi-modal interactions. C: Integration with external plugins like loading objects from Sketchfab to create high-fidelity scenes and special skills like generating animations. D: Prompting edits of existing VR scenes like changing the color of the objects. E: Automated generation of instructional guides and Questioning and Answering about the scene. F: The framework is compatible across platforms and supports the integration of external sensor data.}
      \Description{Examples of diverse use cases and functionalities enabled by Large Language Model for Mixed Reality. This composite image displays six panels, each representing different functionalities of a Mixed Reality framework. A: World Creation - A virtual 3D kitchen constructed with basic geometric shapes such as cubes and rectangles, representing furniture and appliances. B: Multi-modal Interaction - A virtual reality controller with a blue line-rendering trailing from its tip, showing dynamic drawing in a 3D space. C: External Plugin Integration - A virtual room with a wall clock, a picture frame, and furniture, illustrating the incorporation of detailed objects from an external source. D: Edit Existing Worlds - A kitchen scene with a high level of detail, including cabinets, appliances, and decor, demonstrating the ability to modify pre-existing virtual spaces. E: Question and Answer with Scene - A partially disassembled virtual espresso machine with a controller pointing at it, indicating an interactive environment where users can query and receive information. F: Cross-platform and Sensor Integration - A person's wrist wearing a sensor device, with a virtual whale in front of them, suggesting synchronization between physical movements or biometrics and virtual reality actions. Each panel is labeled with a letter from A to F and includes a brief description of the feature or task demonstrated, such as creating environments from scratch, integrating complex objects, or editing existing virtual spaces for accessibility.}
      \label{fig:teaser}
  \end{teaserfigure}

\maketitle

\section{Introduction}

Creating 3D virtual worlds is a challenging task that requires both artistic and technical skills. In addition, 3D content often becomes deprecated and has limited interoperability due to platform and device upgrades. Recently, generative AI models have made considerable progress in producing meshes for objects and scenes \cite{jun2023shap, li20223Ddesigner, lin2023magic3D, singer2023text, text2room, hong20233D, sra2017oasis}. However, few works have ventured beyond visual appearances to bring e.g., interactive and behavioral elements into the generated content. In addition,  existing rendering-based methods require substantial compute and time to generate and render 3D objects, while the quality and resolution of these generations are limited \cite{gao2022nerf, poole2022dreamfusion}.

On the other hand, the rapid advancement in Large Language Models (LLM) like GPT has shown promise in code generation and reasoning \cite{program_synthesis_LLM, code_gen, pangu_coder, web_program_synthesis, jigsaw_code_synthesis}.  An integration of LLMs with a game engine, like Unity \cite{unity}, can enable faster 3D content development and spontaneous user creation,  a core element of mixed reality since its inception. In addition, the 3D mixed reality worlds offer rich, spatial, multimodal information (most are post-symbolic or beyond language) that can potentially help LLMs to better situate their reasoning in the reality that humans live in.

This paper presents \framework  (Large Language Models for Mixed Reality), a framework that enables real-time creation and modification of interactive 3D scenes. \framework \ can create objects that are rich in both visual and behavioral aspects, or make spontaneous and bespoke edits on an existing environment. For example, we leverage LLMR to spawn interactive tools that are self-contained units designed to perform specific functions in virtual and mixed-reality environments. They can be combined to form more complex interactive systems, extending the range and depth of user and AI-driven experiences. These configurations can be saved and transferred across various environments, serving as the building blocks for versatile interactive experiences.

\framework \ is an orchestration of an ensemble of specialized GPTs. At its center is the \textit{Builder} GPT serving as an architect of C\# Unity code for crafting interactive scenes. However, the multitude of tasks falling under virtual world creation renders a standalone coder insufficient. For instance, the ability to meaningfully modify an existing virtual world necessitates a profound semantic understanding of the scene. As humans, we have the ability to infer the properties of objects in the world and can refer to objects in the environment using demonstratives. To simulate the benefits of perceptual access, we incorporated the \textit{Scene Analyzer} GPT. It generates a comprehensive summary of scene objects, offering detailed information when requested, including aspects like size, color, and the functionalities of interactive tools previously generated by \framework. We also implemented the \textit{Skill Library} GPT that determines the relevant skills that are needed for the \textit{Builder} to accomplish the user’s request. In addition, we have observed that the code generated by the \textit{Builder} lacks robustness and frequently contains bugs. To remedy this, we introduce the \textit{Inspector} GPT, which evaluates the \textit{Builder}'s code against a predefined set of rules. This evaluation acts as a protective measure against compilation and run-time errors before the code is executed via the \textit{Compiler} in the Unity Game Engine.

To illustrate the efficacy of our framework in the creation and editing of virtual scenes, we tested LLMR on two sets of 150 prompts encompassing a wide array of creation and modification tasks. Our findings demonstrate LLMR's superior performance in contrast to general-purpose LLMs while emphasizing the performance gain achieved with the addition of each module in our pipeline. In particular, LLMR exhibits 4x reduction in code errors in both an empty and an existing scene, when compared to off-the-shelf GPT-4 \cite{GPT-4}. In the meantime, LLMR can successfully complete sequences of tasks with varying complexities, while keeping the completion time around a minute. These outcomes underscore LLMR's capacity to execute user instructions in real time with a higher degree of robustness.

To evaluate if our framework can generate not only functional code but also interactive worlds that meet users' instructions, we evaluated \framework \ with 11 participants with varying Unity experiences. At a high level, participants found \framework \ to be intuitive and easy to use, and they were able to iteratively achieve desired outputs without much manual scripting. While the framework has limitations such as its unpredictability due to generative models' stochastic nature, and thus is not applicable for all contexts (especially ones that require precise and specific control),  the output generated by \framework \ serves as a starting point for more complex scene generation.

Our paper is organized as follows: we begin by describing prior work and approaches to generating 3D objects and environments for mixed reality in Section 2. In Section 3, we first provide an overview of LLMR followed by details of the function of each module of our framework.  We then discuss important extensions of our framework, such as incorporating plugins, memory management, and cross-platform compatibility, in Sections 4,5,6, respectively. We then present a series of exemplar applications in Section 7 to illustrate the wide range of creations enabled by LLMR. Section 8 Numerical Study includes a comprehensive evaluation of our framework against our design goals: high completion rate, real-time execution, robust against complex tasks, and iterative fine-tuning ability. We follow the Numerical Study with Section 9 User study that evaluates the quality of LLMR's output and presents usability feedback. Finally,  in Section 10, we discuss the limitations and future work for others to build upon.

In summary, our main contributions are the following:
\begin{enumerate}
    \item We introduced a versatile framework for real-time generation of interactive 3D objects and scenes using LLM modules, designed for easy setup with an OpenAI API key and adaptable across various mixed reality tools, environments, and devices.

    \item We carried out extensive evaluations, including a technical ablation study to gauge the framework's performance and reliability, and a user study to derive design recommendations for optimizing the user experience.

    \item We showcased the expanded capabilities of GPT beyond text inputs, illuminating the broader potential of LLM applications, and demonstrated the framework's broad applicability in domains such as remote training, creativity, and accessibility.

     \item We advocate for the interoperability and longevity of mixed reality applications enabled by AI, and thus we openly share the installation package, code, and prompts used in our application and evaluation so that future work can build on top of our framework.
    
\end{enumerate}

\section{Related Work}

Our research on the creation and modification of interactive 3D scenes using natural language is situated at the intersection of large language models (LLMs) and 3D content generation. This section provides an overview of the related work in these areas, highlighting how our work builds upon and extends existing research.

\subsection{Generative 3D Assets}

The generation of 3D assets has been a significant focus in recent research. The work of Li et al. with 3DDesigner \cite{li20223Ddesigner}, Jun and Nichol with Shap·E \cite{jun2023shap}, and Poole et al. with DreamFusion \cite{poole2022dreamfusion} have demonstrated the potential of text guidance and generative models in creating complex and diverse 3D objects. Lin et al. introduce Magic3D \cite{lin2023magic3D}, a high-resolution text-to-3D content creation framework that addresses the limitations of slow optimization and low-resolution output inherent in existing methods like DreamFusion. Recently, Holodiffusion by Karnewar et al. \cite{karnewar2023holodiffusion} furthered the conversation by employing diffusion models for 3D generative modeling. The Instruct-NeRF2NeRF method \cite{haque2023instruct} and advancements like Pointclip v2 \cite{zhu2023pointclip} as well as the work of Roberts et al. \cite{roberts2022steps} have explored the power of prompting techniques in 3D open-world learning. A comprehensive review of Neural Radiance Field (NeRF) models by Gao et al. \cite{gao2022nerf} adds to our understanding of this rapidly growing field and aligns with our approach of enabling LLMs to interpret non-linguistic or non-symbolic information. Our approach extends beyond visual appearances to incorporate interactive and behavioral elements into the generated content.

\subsection{Generative Interactive 3D Environments}

In addition to generating objects, the creation of interactive 3D environments has been further explored, with contributions from Wang et al. with Voyager \cite{wang2023voyager}, Singer et al. with MAV3D \cite{singer2023text}, and Höllein, Lukas, et al. with Text2Room \cite{text2room}. Volum et al. has shown that LLMs can be used to guide NPC interactions with a virtual environment \cite{ironsword}. Wang et al. also introduced Chat-3D \cite{wang2023chat}, a system that focuses on universal dialogues for 3D scenes, which is further augmented by the work of Hong et al. with 3D-LLM \cite{hong20233D}. New approaches like Oasis \cite{sra2017oasis} and Procedurally Generated Virtual Reality \cite{sra2016procedurally} add novel perspectives. Recent advancements such as Interactive Example-Based Terrain Authoring with Conditional Generative Adversarial Networks by Guérin et al. \cite{guerin2017interactive} add a layer of complexity to how terrains can be generated from simple user inputs. Research by Freiknecht and Effelsberg \cite{freiknecht2017survey}, Cao et al. \cite{cao2023hexplane}, and Song et al. \cite{song2023objectstitch} has focused on the balance between realism and algorithmic performance. DeepSpace introduced a novel method of mood-based texture generation from music \cite{sra2017deepspace}, adding another layer of complexity to asset generation.  While these contributions are significant in building interactive 3D spaces, the interplay between AI and mixed reality in these environments remains an open question. Our work tackles this gap by bringing the capabilities of LLMs to a real-time Unity editor for Mixed Reality applications.

\subsection{Editor Support for Mixed Reality Development}

Mixed Reality (XR) development has been explored by Hirzle et al. \cite{hirzle2023xr} and Fidalgo et al. \cite{fidalgo2023survey}, who provide comprehensive reviews at the intersection of AI and XR. Lindlbauer et al. \cite{lindlbauer2019context} and Cheng et al. \cite{cheng2021semanticadapt} focus on the automatic adaptation of MR interfaces, a line of work that is relevant for multi-user XR experiences, as shown by Mandi et al. with RoCo \cite{mandi2023roco}. Thoravi Kumaravel et al. \cite{thoravi2019loki} complement these efforts by focusing on bi-directional mixed-reality telepresence. Compared with prior work, we allow users to directly authorize the environment using natural language.

\subsection{LLMs Interpreting Spatial, Non-Linguistic Information}

Lastly, many have pushed the boundary of LLMs by inputting non-linguistic information (which was not in the training set), such as for visual programming \cite{zhou2023instructpipe} or processing sensor data \cite{liu2023large}. More related to our work is using LLM to interface with spatial, embodied data. Work of Zhang et al. with MotionGPT \cite{zhang2023motiongpt}, Wu et al.'s work on Embodied Task Planning \cite{wu2023embodied} as well as Richardson et al. with TEXTure \cite{richardson2023texture}. Daffara et al. \cite{daffara2020authorive} and Rana et al. \cite{rana2023sayplan} further extended these concepts to include demonstrations and task planning. Driess et al. with PaLM-E \cite{driess2023palm} has shown the potential of LLMs in generating human motion, texturing 3D shapes, and incorporating real-world sensor modalities, respectively. These efforts are complemented by Xu et al. with XAIR \cite{xu2023xair}, which focuses on explainable AI in augmented reality. We hope LLMR contribute to the improvement of LLM's capability of spatial reasoning and world understanding.

\begin{figure*} [hbt!]
    \centering
    \includegraphics[width=\textwidth]{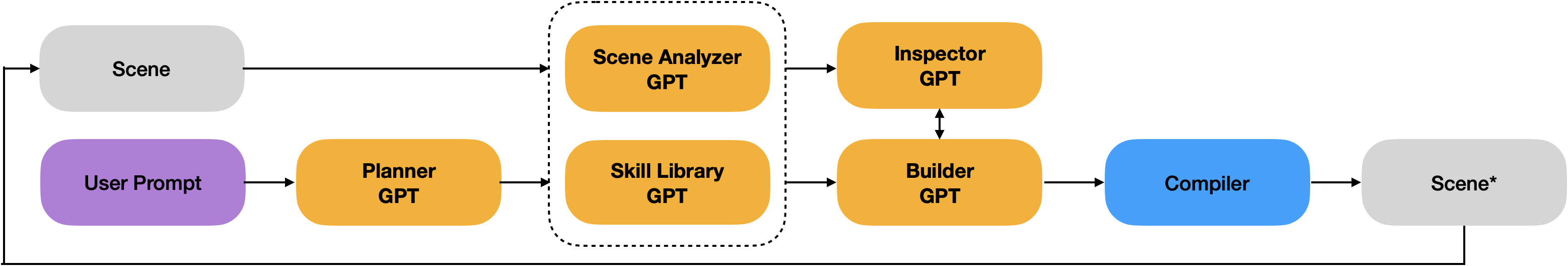}
    \caption{\textit{Large Language Model for Mixed Reality (\textit{\framework}) architecture for real-time interactive 3D scene generation}. Starting from the left, a user prompt and the existing 3D scene ($\Omega$) are fed into the \textit{Planner} (P) and \textit{Scene Analyzer} (SA) modules, respectively. The \textit{Planner} decomposes the user prompt into a sequence of sub-prompts, while the \textit{SA} summarizes the current scene elements. These are then integrated with a \textit{Skill Library} (SL) to guide the \textit{Builder} (B) module, which generates the appropriate code. The \textit{Inspector} (I) module iteratively checks the generated code for compilation and run-time errors. Upon receiving the green light from the \textit{Inspector}, the code is compiled using the Roslyn Compiler and executed in the Unity Engine to produce the desired 3D scene and functionalities as specified by the user.}
    \Description{Large Language Model for Mixed Reality architecture for real-time interactive 3D scene generation. This flowchart illustrates the architecture for a Large Language Model framework used in real-time interactive 3D scene generation. Starting from the left, a user's input prompt, shown in purple, and an existing 3D scene, shown in gray, initiate the process. These inputs feed into two modules: the Planner, depicted in purple, breaks down the user's prompt into sub-tasks; the Scene Analyzer, in orange, outlines current scene elements. Both utilize the GPT model for processing. The outputs from these modules converge with a Skill Library, represented in orange and powered by GPT, which informs the Builder module, also in orange. The Builder generates code that is checked by the Inspector module, in orange, for errors in a feedback loop. Upon approval, the code is compiled, signified by a blue block, and executed in the Unity Engine to produce an updated 3D scene, indicated at the end of the process by a gray block labeled 'Scene*'. This architecture demonstrates a complex, iterative design for modifying and generating 3D environments in response to user commands.}
    \Description{Large Language Model for Mixed Reality architecture for real-time interactive 3D scene generation. This flowchart illustrates the architecture for a Large Language Model framework used in real-time interactive 3D scene generation. Starting from the left, a user's input prompt, shown in purple, and an existing 3D scene, shown in gray, initiate the process. These inputs feed into two modules: the Planner, depicted in purple, breaks down the user's prompt into sub-tasks; the Scene Analyzer, in orange, outlines current scene elements. Both utilize the GPT model for processing. The outputs from these modules converge with a Skill Library, represented in orange and powered by GPT, which informs the Builder module, also in orange. The Builder generates code that is checked by the Inspector module, in orange, for errors in a feedback loop. Upon approval, the code is compiled, signified by a blue block, and executed in the Unity Engine to produce an updated 3D scene, indicated at the end of the process by a gray block labeled 'Scene*'. This architecture demonstrates a complex, iterative design for modifying and generating 3D environments in response to user commands.}
    \label{fig:framework}
\end{figure*}

\section{\framework: a framework for generating real-time, interactive 3D worlds using large language models}

Large language models are capable code generators, and their ability to synthesize programs has been extensively tested \cite{program_synthesis_LLM, code_gen, pangu_coder, web_program_synthesis, jigsaw_code_synthesis}. Scripting in a game engine, however, is especially challenging given the multitude of tasks and the complexity of the development environment. For a non-comprehensive list, generating a realistic 3D world may involve object creation, texturing, behavior programming, event scripting, animations, particle effects, lighting, and user interface \cite{game_design_book}. Prompting these elements in real time requires a framework that understands the virtual scene, interprets user intention, and generates high-quality code. To this end, we present Large Language Model for Mixed Reality (\textit{\framework}), a framework that enables real-time creation and modification of interactive 3D scenes using natural language.

\framework \ is an orchestration of language models, each contextualized with a distinct metaprompt to outline its role, as illustrated in Figure \ref{fig:framework} and Algorithm \ref{LLMR_algo}. A metaprompt is a specially crafted input sequence or context that guides an LLM's behavior or output, enabling more focused or nuanced responses than standard prompts. We start with the \textit{Planner}, which breaks down the user's request into a sequence of appropriately scoped instructions. These instructions, along with a concise summary of the existing scene from the \textit{Scene Analyzer} and extra knowledge for specialized skills from the \textit{Skill Library}, are used as inputs to the central module called \textit{Builder}, which generates code to fulfill these instructions. In addition, we use a separate \textit{Inspector} module to check the \textit{Builder}'s generated code against potential compilation and run-time errors before finally executing the code.

The task of generating interactive 3D scenes boils down to generating and executing appropriate code snippets to accomplish the user's prompt. Formally, denote the user's request by $u$ and the current 3D world by $\Omega$ (which may be empty), we wish to draw sample $x \sim \mathcal{P}(x|u,\Omega)$, where $\mathcal{P}$ is the distribution of syntactically valid, request-fulfilling code. We then compile and execute $x$ at run-time under the Unity Engine \cite{unity}, a development platform for creating virtual scenes that suits our needs. Below, we detail each module and explain the design choices that enable various aspects of prompting a virtual world into existence.

        

\begin{algorithm}
\caption{LLMR}
\label{LLMR_algo}
\Input{$u$: user's request; $\Omega$: current scene}
\Require{$\mathrm{A}(s|u,\Omega)$: Scene Analyzer; \\
$\mathrm{P}(u_1, ..., u_N | u,s)$: Planner; \\
$\mathrm{L}(h|u)$: Skill Library; \\
$\mathrm{B}(x|u,s,h)$: Builder; \\
$\mathrm{I}(r,v|u,s,x)$: Inspector.}

$s \sim \mathrm{A}(\cdot | u,\Omega) $\;
$(u_1, ..., u_N ) \sim \mathrm{P}(\cdot | u,s) $ \Comment{Decomposes the request into suitable instructions.}\
$\Omega_1 \gets \Omega$\;
\For{$i=1:N$}{
$s_i \sim \mathrm{A}( \cdot | u_i,\Omega_i)$  \Comment{Analyze the current scene.}
$h_i \sim \mathrm{L}( \cdot | u_i)$ \Comment{Retrieve required skills.}
$x_i \gets \textrm{GenerateCodeWithInspection}(u_i,s_i,h_i)$\;
$\Omega_{i+1} \gets \textrm{CompileAndRun}(x_i, \Omega_i)$\;
}
\end{algorithm}

\begin{figure*}[hbt!]
\centering
  \includegraphics[width=.95\textwidth]{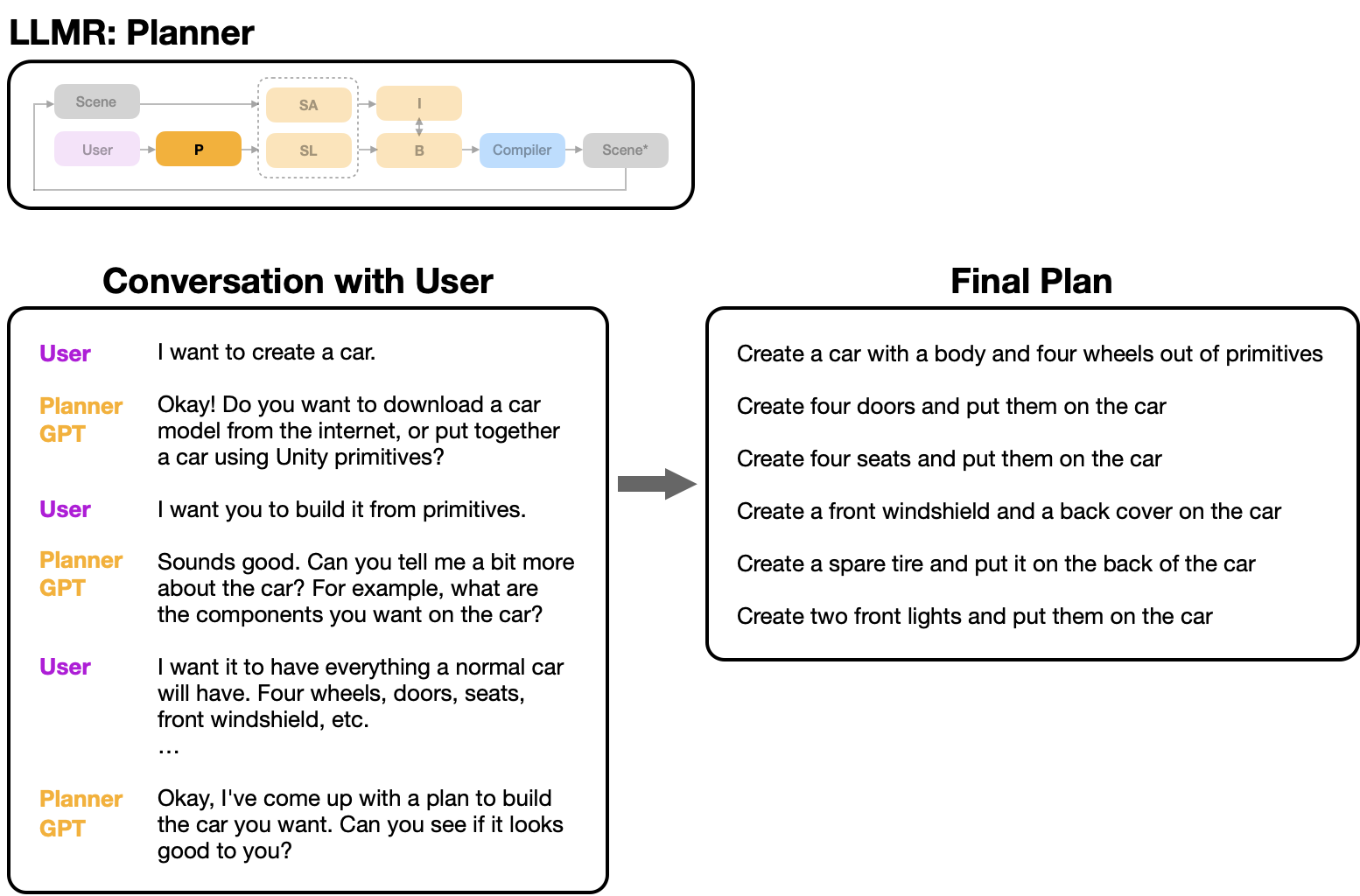}
\caption{The \textit{Planner and its role in breaking down a user's high-level request into a sequence of manageable subtasks $(u_1, u_2, \ldots, u_n)$}. The \textit{Planner} engages in a user-oriented conversation to determine the appropriate scope and granularity of each subtask. Following this, the \textit{Builder} executes the plan by generating code $(x_1, x_2, \ldots x_n)$ for each subtask, effectively carrying out the user's initial request.}
\Description{Planner and its role in breaking down a user's high-level request into a sequence of manageable subtasks. This image is split into two main sections and an inset at the top. The inset is a flowchart featuring the components from the previous figure: Scene, User, Planner (P), Scene Analyzer (SA), Skill Library (SL), Builder (B), Compiler, and the resulting Scene*. Below this, the left section is titled "Conversation with User" and displays a dialogue between a user and the Planner GPT. The user expresses a desire to create a car from primitives rather than downloading a model. The Planner GPT inquires about specific components of the car to clarify the request. On the right, the "Final Plan" section lists a sequence of tasks derived from the conversation: creating a car body with wheels, doors, seats, windshields, a spare tire, and front lights, all from primitives. This demonstrates the Planner's role in translating a user's high-level concept into actionable subtasks for the Builder to execute in code, following the user's initial request for a car built from basic elements.}
\Description{Planner and its role in breaking down a user's high-level request into a sequence of manageable subtasks. This image is split into two main sections and an inset at the top. The inset is a flowchart featuring the components from the previous figure: Scene, User, Planner (P), Scene Analyzer (SA), Skill Library (SL), Builder (B), Compiler, and the resulting Scene*. Below this, the left section is titled "Conversation with User" and displays a dialogue between a user and the Planner GPT. The user expresses a desire to create a car from primitives rather than downloading a model. The Planner GPT inquires about specific components of the car to clarify the request. On the right, the "Final Plan" section lists a sequence of tasks derived from the conversation: creating a car body with wheels, doors, seats, windshields, a spare tire, and front lights, all from primitives. This demonstrates the Planner's role in translating a user's high-level concept into actionable subtasks for the Builder to execute in code, following the user's initial request for a car built from basic elements.}
\label{fig:Planner}
\end{figure*}

\newpage
\subsection{Planner}

Prompting a world into existence can be a hefty task. "Create a city and all its denizens" is a valid request, albeit one that is overly ambitious to achieve in a single step. Following the common wisdom "nothing is particularly hard if broken into small jobs", instead of directly sampling from $\mathcal{P}(x|,u, \Omega)$, we propose a \textit{Planner} $P: u \mapsto (u_1, u_2, ..., u_N)$ to decompose each prompt into subtasks within an appropriate scope, then use autoregressive sampling to carry out these subtasks via a sequence of generated code $(x_1, x_2, ..., x_N)$:
\begin{equation}
\begin{split}
    &\mathcal{P}(x_1, x_2, ..., x_N | u_1, u_2, ..., u_N, \Omega) = \mathcal{P}(x_1|u_1, u_2, ..., u_N,\Omega) \ \times \\  &\times \prod_{n=1}^{N-1} \mathcal{P}(x_{n+1} | \bm{x}^n, u_1, u_2, ..., u_N, \Omega) \\
    &= \mathcal{P}(x_1|u_1,\Omega) \prod_{n=1}^{N-1} \mathcal{P}(x_{n+1} | \bm{x}^n, u_{n+1}, \Omega)
\end{split}
\end{equation}

where $\bm{x}^n \coloneqq  (x_1, ..., x_n)$. The second quality follows by assuming independence of code generations and requests at different steps, $x_n \ind u_m, \forall m\ne n$. An illustration for this procedure is provided in Figure \ref{fig:Planner}. However, sampling from $\mathcal{P}(x_{n+1} | \bm{x}^n, u_{n+1}, \Omega)$ may be difficult for a language model, because it has to \textit{infer} the effect of $(x_1, ..., x_n)$ on the initial world $\Omega$ before writing code $x_{n+1}$. To remove the guesswork, we leverage a runtime compiler $R$ to execute $(x_1, ... x_n)$ in order, each time getting a new world state $\Omega_{n+1} = R(x_{n+1},\Omega_n)$. We can then rewrite:
\begin{equation}
\begin{split}
\mathcal{P}(x_{n+1} | \bm{x}^n, u_{n+1}, \Omega) = \mathcal{P}(x_{n+1} | x_n, u_{n+1}, \Omega_n),
\end{split}
\end{equation}

where we assume $\{x_i\}_{i=1}^n$ is Markovian when conditioned on $\Omega_n$. That is, the current world state is rich enough to capture all previous executions past the most recent one.

In principle, it is possible for the user to limit their prompts within a certain difficulty so that the decomposition is unnecessary. However, the user may not know the appropriate task scope a priori (if creating a city is too hard, how about a single house? Or a room in the house?) As a result, having a properly configured \textit{Planner} makes the framework robust to prompts of varying difficulty. In addition, the user may have different levels of details in their prompt. For example, "Creating a car" is a valid request that nevertheless does not specify its appearance or functionality. Here, the \textit{Planner} serves as a conversational assistant that interacts with the user to devise a plan with an appropriate scope and granularity, which significantly improves the user experience.

\begin{figure*}[hbt!]
\centering
     \includegraphics[width=\textwidth]{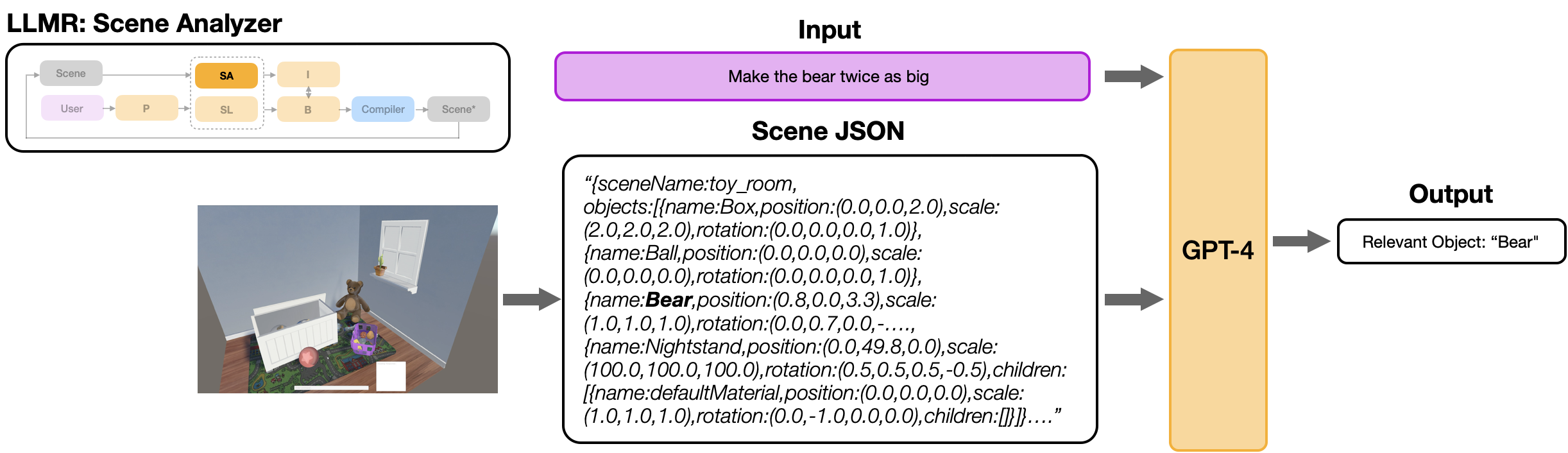}
\caption{\textit{Scene Analyzer module. }The virtual scene, depicted in the bottom-left corner, is converted into a parsed scene hierarchy in JSON format. This, along with the user request, serves as input to the \textit{Scene Analyzer}. The output is a filtered, relevant summary of the scene, which is then used for conditioning subsequent modules like the \textit{Builder}. The process optimizes the utilization of the language model's fixed context window and enhances focus on objects relevant to the user prompt.}
\Description{Scene Analyzer module. This diagram illustrates the process flow of the Scene Analyzer module within a Large Language Model for Mixed Reality system. The flow begins with a virtual scene depicted in the bottom-left, showing a room with various objects such as a bear, a ball, and furniture. This visual scene is translated into a structured text format known as JSON, capturing the properties of each object like name, position, rotation, and scale. A user input, highlighted in a purple box with the text "Make the bear twice as big", feeds into the Scene Analyzer. The Analyzer processes this input along with the scene's JSON representation to produce an output that identifies and prioritizes elements relevant to the user's request. Here, the output is indicated by a goldenrod-colored box pointing to the term "Relevant Object: 'Bear'" after being processed by GPT-4, suggesting that the Scene Analyzer has successfully isolated the bear object as the focus for further action by subsequent system modules such as the Builder. This step is crucial for optimizing the language model's context window and ensuring focus on user-relevant scene elements.}
\label{fig:Scene_Analyzer}
\end{figure*}

\newpage
\subsection{Scene Analyzer}\label{sec:SA}

There are many possible representations of a virtual world $\Omega$ that may include visual, behavioral, and auditory elements. In this work, we derive $\Omega$ from the Unity scene hierarchy, which contains all existing game objects, their attached components, and their parent-child relations. The hierarchy is parsed into a JSON string and can then be used as input to language models. However, directly using the raw JSON string as input proves to be infeasible in practice. First, most prompts only require interactions with a small subset of $\Omega$, so it is unnecessary and even distracting to use its entirety as input. Second, LLMs have a fixed context window $W$ that serves as its short-term memory, which has to contain its metaprompt, few-shot examples, user prompt, and generative output \cite{llm_survey}. For example, GPT-4 supports either 8k or 32k tokens for maximum number of token at a time\cite{GPT-4}, but even the 32k token limit can be insufficient, particularly for intricate scenes containing numerous objects, each consisting of multiple components.

To tackle these issues, we created a separate module termed the \textit{Scene 
 Analyzer}, which is a properly prompted LLM $\mathrm{A}(s|u,\Omega)$ that outputs a succinct summary of $\Omega$ conditional on the user request. At a high level, one can think of the \textit{Scene Analyzer} as a means of perception that relays an abstraction of the environment for downstream processing. An illustration of the module is provided in Figure \ref{fig:Scene_Analyzer}. Concretely, the output $s_n \sim \mathrm{A}(\cdot |u, \Omega_n)$ is used to reparametrize the density at each sampling step:
\begin{equation}
\begin{split}
 \mathcal{P}(x_{n+1} | x_n, u_{n+1}, s_n) \sim \mathcal{P}(x_{n+1} | x_n, u_{n+1}, \Omega_n)
\end{split}
\end{equation}

\begin{figure*}[hbt!]
\centering
     \includegraphics[width=\textwidth]{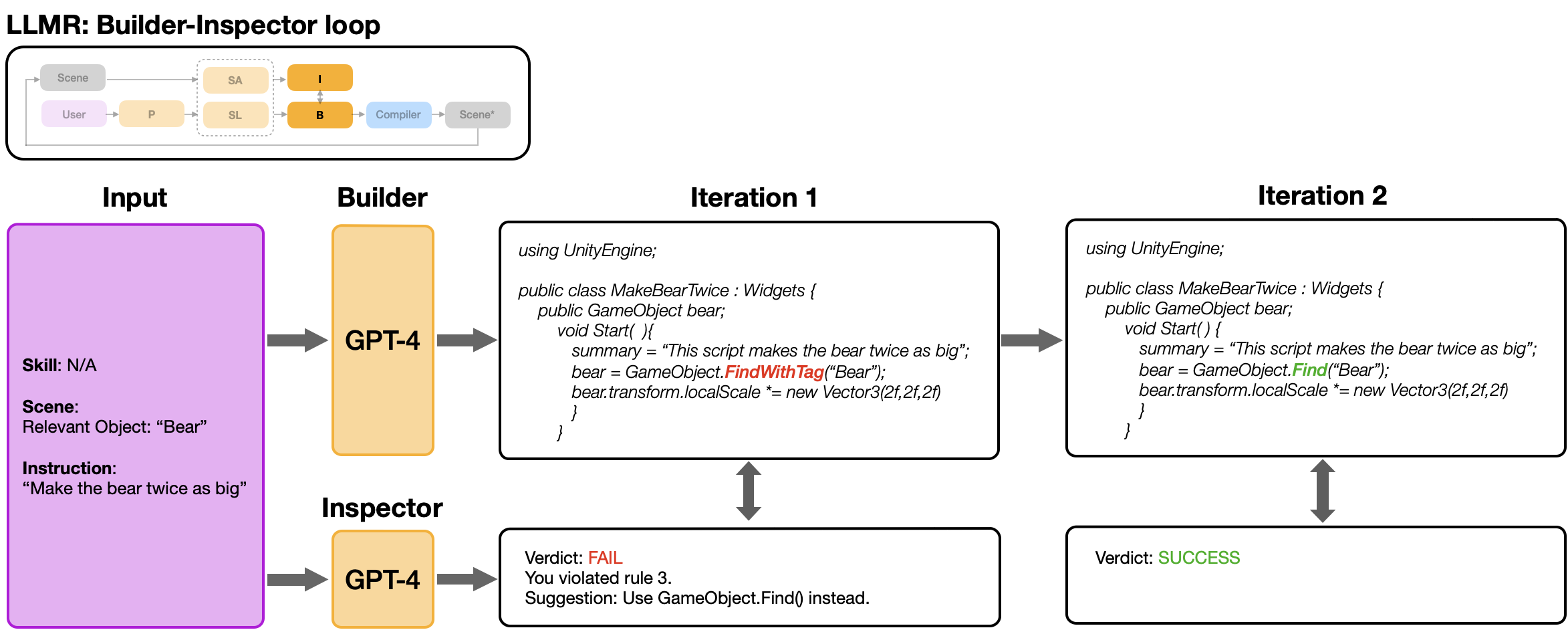}
\caption{\textit{Builder-Inspector paradigm in \framework. }The \textit{Builder} module $\mathrm{B}(x | u, s)$ generates code based on user input and current state. The generated code is then inspected by the \textit{Inspector} module $\mathrm{I}(r,v|x,s)$ for compilation and run-time errors. If errors are found, indicated by verdict $v$, the \textit{Inspector} provides suggestions $r$ for corrections. The process iterates until either the code passes inspection or a maximum number of inspections $T$ is reached. This feedback loop significantly enhances the quality of the generated scripts.}
\Description{Builder-Inspector paradigm. The image is a diagram of the "Builder-Inspector loop" concept in a framework. The diagram is divided into several sections, including "Input," "Builder," and two iterations, "Iteration 1" and "Iteration 2," with an "Inspector" at the bottom. The Input section at the top left shows a UI interface with various buttons and a field to enter a command, labeled "Skill: NA, Scene: Object 'Box', Instruction: 'Make the box twice as big'". The Builder section has an orange background and is connected to Iteration 1, which contains a code snippet with a comment "// command: 'Make the box twice as big'", and a line of code that incorrectly scales a "Box" object. The Inspector, which has a purple background, evaluates the code from Iteration 1, providing a "Verdict: FAIL" and a suggestion to use "GameObject.Find()" instead. Iteration 2 shows a revised code snippet with the suggested correction. The Inspector then gives a "Verdict: SUCCESS". Arrows indicate the flow from Input to Builder, then to each iteration, and feedback from the Inspector. This diagram illustrates a feedback loop process where code is generated, inspected, and iterated upon until it passes inspection or reaches a maximum number of iterations, enhancing the quality of generated scripts.}
\label{fig:Inspector}
\end{figure*}

\newpage
\subsection{Builder-Inspector}
Central to \framework \ is the \textit{Builder} $\mathrm{B}(x | u, s)$, a module responsible for generating code conditional on the user prompt. It serves as our main apparatus for approximating $\mathcal{P}$. In other words, we hope
\begin{equation}
\begin{split}
\mathrm{B}(x | x_n, u_{n+1}, s_n) \approx \mathcal{P}(x_{n+1} | x_n, u_{n+1}, s_n),
\end{split}
\end{equation}
holds with a carefully crafted metaprompt and enough in-context demonstrations. In practice, however, the complex nature of creating a virtual world makes the approximation unsatisfactory even with as many examples as the context length allows. 
This is largely because the \textit{Builder} module is asked to accomplish the instructions with some creativity while faithfully following an extensive list of specific guidelines that align the output, which causes to \textit{Builder} to have a "cognitive overload".

\begin{algorithm}[hbt!]
\caption{Generate Code With Inspection}
\label{Builder_Inspector_algo}
\Input{$u$: user's request, $s$: scene summary, $h$: additional hint.}
\Require{$\mathrm{B}(x|u,s,h)$: Builder; \\
$\mathrm{I}(r,v|u,s,x)$: Inspector; \\
T: maximum number of inspections.}

$t \gets 0$\;
$r_0 \gets \emptyset$\;
$v_0 \gets $ False\;
\While {$ t<T$ and $v_t$ is false}{
$x_t \sim \mathrm{B}(\cdot | u,s,h,r_t)$  \Comment{Builder writes code $x_t$}
$(r_t,v_t) \sim \mathrm{I}(\cdot | s,x)$  \Comment{Inspector checks code, outputs verdict $v_t$ and suggestion $r_t$}
$t \gets t+1$\;
}
\Return x
\end{algorithm}

\newpage
To ameliorate this, we introduce another module, the \textit{Inspector} $\mathrm{I}(r,v|x,s)$, that checks the \textit{Builder}'s generated code for compilation and run-time errors. In the case of a failed inspection indicated by verdict $v$, the \textit{Inspector} outputs a suggestion $r$ for potential fixes and prompts the \textit{Builder} to make another attempt. As a result, the \textit{Builder} and \textit{Inspector} work in tandem to write and self-debug code, forming a feedback system that significantly improves the quality of the generated scripts. We outline this paradigm in Algorithm \ref{Builder_Inspector_algo} and illustrate it in Figure \ref{fig:Inspector}. Interestingly, the \textit{Inspector} excels at catching errors even if the same guidelines in its metaprompt are present in the \textit{Builder}. One possibility is that this is due to providing a more extensive list of negative and positive examples to the \textit{Inspector}. Still, when the \textit{Builder} is provided with the same examples, performance is not as high. Our intuition for this is that verifying a snippet of code is easier than writing the said code, or the two tasks bear different failure modes that can be effectively hedged.


\subsection{Compilation, Save and Reload}\label{sec:save_reload}

After the \textit{Builder}-generated script passes the inspection, we follow the approach in \cite{roberts2022steps} to compile and execute the scripts at runtime through the Roslyn C\# compiler \cite{roslyn}. The inclusion of run-time compilation elevates \framework \ from an offline development tool to a real-time generative framework.

To enable iterative design, users can save their generations and selectively reload the saved generations in the existing or new scene without having to repeat the prompting process. The generated output is saved as C\# scripts and reattached to the Compiler to be compiled at runtime.  A one-sentence summary of each script's function is saved, so alternatively, the output can also be regenerated by the framework based on the summary.

\begin{figure*}[hbt!]
    \centering
     \includegraphics[width=\textwidth]{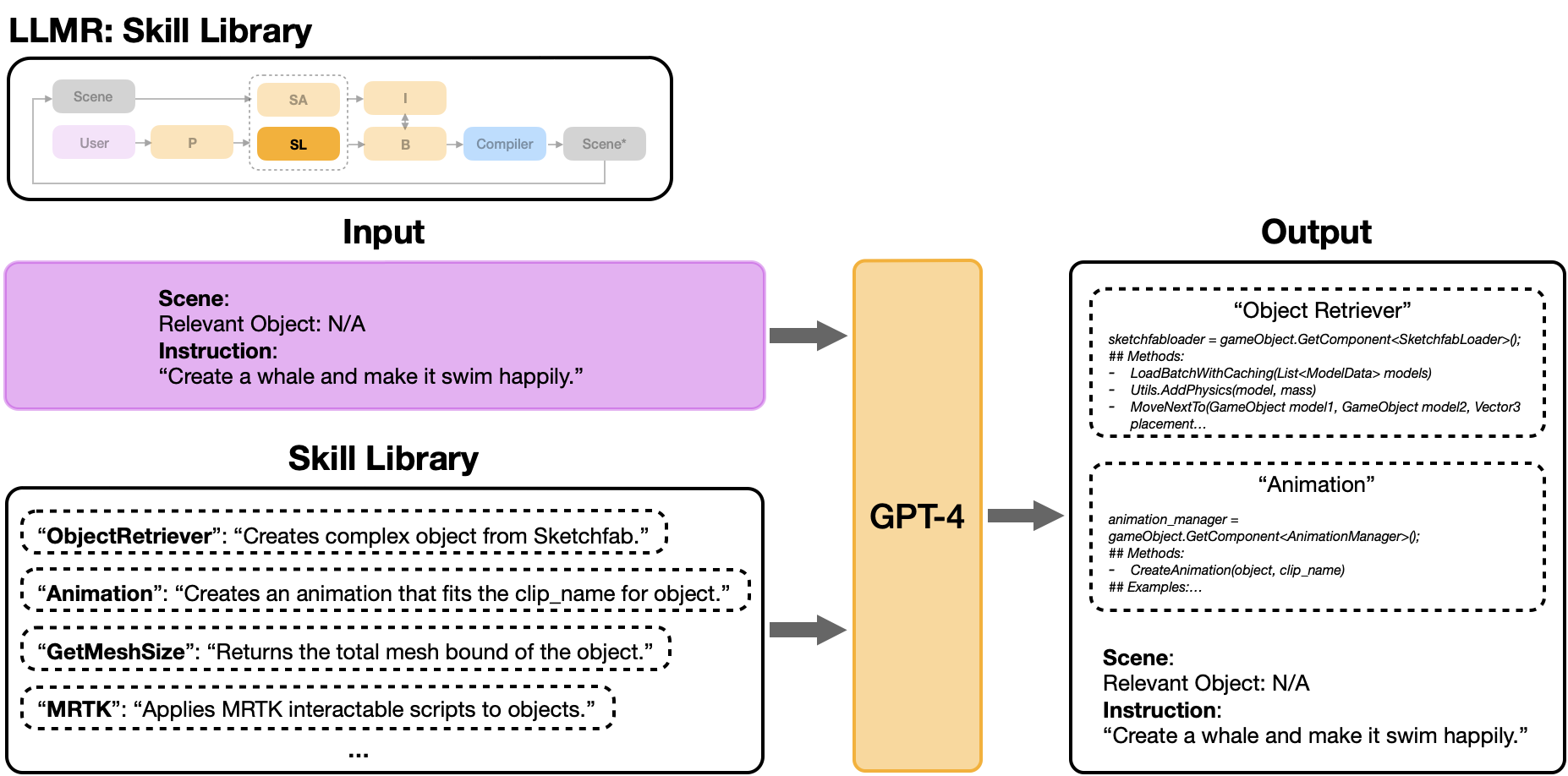}
    \caption{\textit{\textit{Skill Library} module workflow.} On the left, the module receives inputs from the \textit{Scene Analyzer} and a user prompt "create a whale and make it swim happily". A list of skills is provided to the SL GPT module in its metaprompt, which also contains a high-level summary of available skills such as object retrieval and animation. The module then identifies and outputs the most relevant skills (in this case, object retriever and animation) to the \textit{Builder}, which subsequently utilizes these tools for implementation.}
    \Description{Skill Library Workflow. This image depicts the workflow of a 'Skill Library' module in a software context. The diagram is split into two main sections, 'Input' on the left and 'Output' on the right, with an arrow labeled 'GPT-4' pointing from the left section to the right, indicating the flow of processing. The 'Input' section shows a simplified user interface with buttons for 'Scene,' 'SA,' 'User,' 'P,' 'SL,' 'Developer,' and 'Server,' and below is a text box with the user prompt 'create a whale and make it swim happily.' Below this, a list of skills is detailed, including 'ObjectRetriever,' 'Animation,' 'GetMeshSize,' and 'MRTK,' each with a brief description of its function. The 'Output' section contains code snippets and references to 'Object Retriever' and 'Animation,' suggesting these are the selected outputs from the Skill Library module. The bottom of both sections restates the user prompt and mentions that the relevant scene and object are not applicable.}
    \label{fig:skill_lib}
    
\end{figure*}

\subsection{Skill Library}

The creation of the \textit{Skill Library} Module is motivated by two primary challenges. The first is the token size limitation imposed by the GPT architecture on the context, or the "metaprompt," provided to the \textit{Builder}. Typically, the \textit{Builder} is presented with a comprehensive list of various APIs and plug-ins that could be employed to meet the user's needs. As the range of available skills expands, this list lengthens, eventually surpassing GPT's token size limit for public users. The second challenge lies in the \textit{Builder}'s attention capacity, which appears to be limited. Even when we attempt to condense all the available skills into the \textit{Builder}'s metaprompt, it struggles to keep track of a specific skill when the list becomes too lengthy. This limitation is further exacerbated by the necessity to include precise coding examples for each plugin to ensure their effective utilization by GPT. To address these challenges, we created the \textit{Skill Library} module, denoted as $\mathrm{L}(h|u)$, which serves as a centralized repository for all available skills and as an attention mechanism that retrieves only the skills relevant to a specific user prompt. We illustrate this module in Figure \ref{fig:skill_lib}.

Formally, a specialized GPT is provided with a metaprompt containing two essential pieces of information: 1) a high-level summary of the available skills, and 2) the user's prompt. The GPT model is tasked with identifying either a single skill or a subset of skills that are most pertinent to the user's request. The \textit{Skill Library} remains efficient and small in token size because it only needs the high-level descriptions of each skill, while the specific usage details, as well as positive and negative examples, are stored separately. Once the relevant skills are identified, their detailed information and usage examples are fetched and passed on to the \textit{Builder} for implementation.

\begin{equation}
h_i \sim L( \cdot | u_i) \quad \text{(Retrieve required skills, if any.)}
\end{equation}

As an illustrative example, consider a skill we created for GPT's use, which leverages a combination of generative and contrastive models along with the Sketchfab API to source and integrate 3D models into a scene. We have also created skills that allow the generation of animation of a rigged object in real-time \cite{huang2023real}. While we delve into the specifics of a skill in the next section, it is worth noting that the \textit{Skill Library} only receives a high-level summary of how this particular skill functions, along with similar descriptors for other skills. The actual examples needed to use this skill are then retrieved and supplied to the \textit{Builder} for execution. 
 
\begin{equation}
\mathrm{B}(x|u,s,h): \text{Builder}; where h = \text{retrieved skills from} L
\end{equation}

This approach ensures that the \textit{Skill Library} and the \textit{Builder} work in tandem to efficiently and effectively generate code that fulfills the user's request while overcoming the token size and attention capacity limitations of LLMs.

\begin{figure*}
    \centering
     \includegraphics[width=\textwidth]{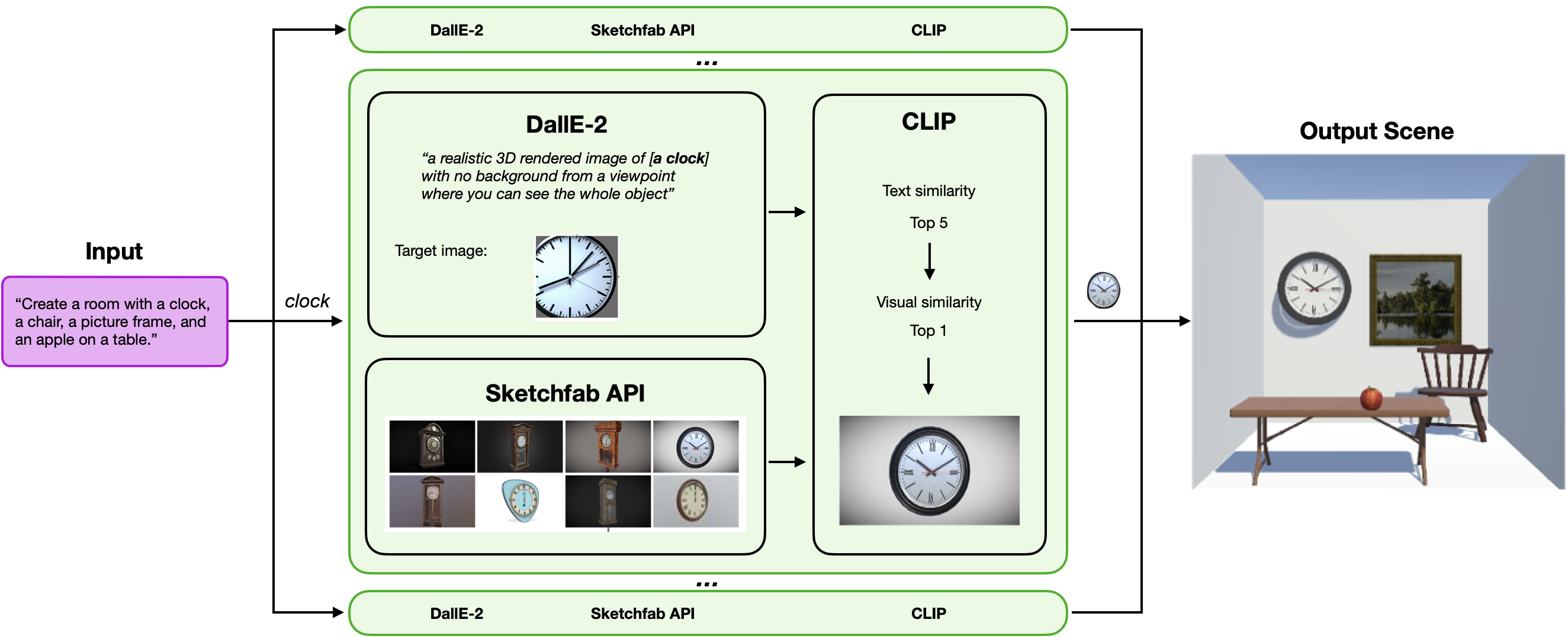}
    \caption{\textit{Object Retriever pipeline for generating a 3D scene}. The user provides a prompt for a scene containing a clock, a picture frame, a chair, and an apple on a table. For each object (e.g., clock), the pipeline uses DALL-E 2 to create a target 3D image. Concurrently, multiple screenshots of potential matches from open-source Sketchfab models are downloaded using the object label as the query. CLIP is employed to generate embeddings for these images which includes the target image. The top 5 candidates in the language similarity space are selected. The final object is then chosen based on the highest visual similarity to the target image. This sequence is repeated for each object in the prompt to assemble the complete 3D scene, as shown on the far right.}
    \Description{Object Retriever pipeline for generating a 3D scene. This diagram depicts the Object Retriever pipeline for generating a 3D scene based on a user prompt. The input stage begins with a textual prompt for a scene with specific objects. For each object, such as a clock, DALL-E 2 generates a target 3D image. Simultaneously, the Sketchfab API retrieves multiple potential matches for each object, displayed as thumbnails. CLIP then processes these images to produce embeddings that include the target image. The top 5 candidates are selected based on text similarity, and the final choice is made based on the highest visual similarity to the target image. This process is repeated for each object in the prompt. The final output is a rendered 3D scene on the right, featuring a wall clock, a picture frame on a wall, a wooden chair, and a table with an apple on it, all placed within a room.}
    \label{fig:object-retrieve}
\end{figure*}

\section{Incorporating Existing Open-Source 3D Assets}

\begin{figure*}[hbt!]
    \centering
     \includegraphics[width=\textwidth]{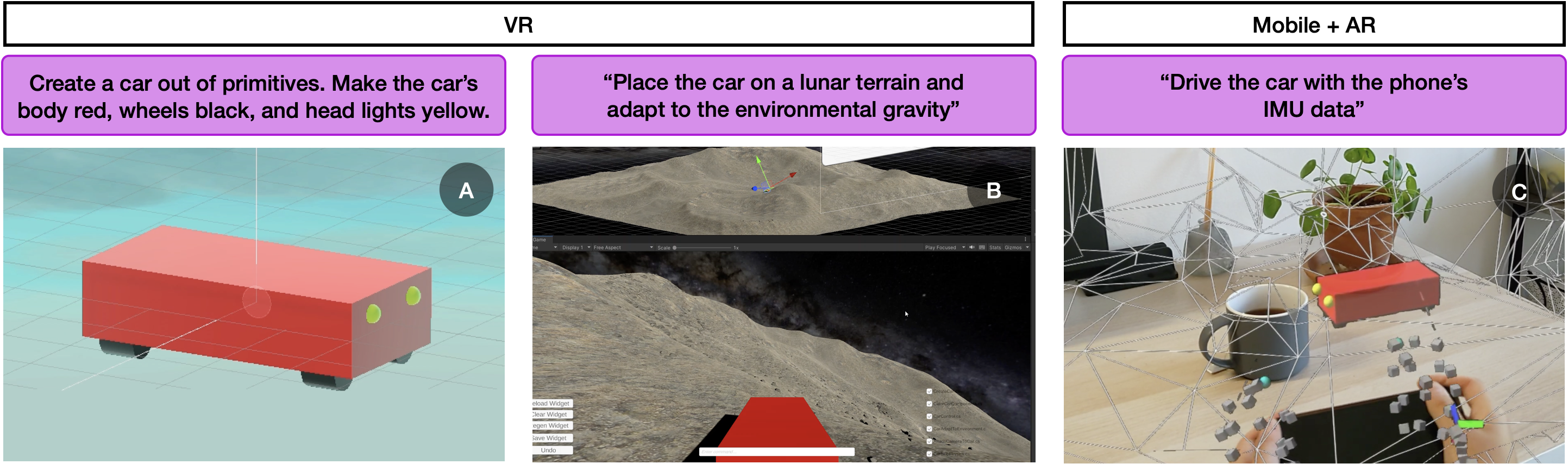}
    \caption{\textit{Cross-Platform and Cross-Scene Transferability made possible by \framework}. The left panel shows a car automatically created by \framework\  using Unity primitives, complete with color and composite features (e.g., wheels and headlights), controllable via keyboard inputs. The middle panel displays the same car transferred to a different Unity scene featuring moon-like gravity and terrain. The right panel showcases the framework's adaptability across platforms by illustrating how the car can collide with objects in the physical world and can be controlled using IMU data from a user's mobile phone.}
    \Description{Cross-Platform and Cross-Scene Transferability made possible by LLMR. This composite image illustrates the transferability of a car model across different platforms and scenes using a specific framework. The first panel on the left, labeled 'A' under the VR category, shows a simple car model created using Unity primitives with a red body, wheels, black, and headlights in yellow, indicating it can be controlled with keyboard inputs. The middle panel, labeled 'B', under the VR category, depicts the same car model placed on a lunar terrain with moon-like gravity within Unity, demonstrating the model's adaptability to different virtual environments. The rightmost panel, labeled 'C' under the Mobile + AR category, showcases the car in a real-world setting overlayed through augmented reality (AR) on a mobile device, showing the car's ability to collide with physical objects and be controlled using the phone's Inertial Measurement Unit (IMU) data.}
    \label{fig:car-crossplatform}
    
\end{figure*}

The process of generating interactive 3D scenes often involves the creation and placement of various objects. For instance, a request to create an office space might be decomposed into the generation of a desk, a chair, a lamp, and a clock. While it is possible to generate these objects using primitives, a method that works well even for composite objects like a car or an entire room (depicted in the car of Figure \ref{fig:car-crossplatform} and the kitchen of Figure \ref{fig:teaser}), there is a need to leverage the intricate objects created by artists and 3D developers that exhibit high real-world fidelity. Previous work has utilized objects from Sketchfab \cite{roberts2022surreal,roberts2022steps} and used the priors of GPT to size them accordingly to the real world. However, this approach encounters challenges when the user prompts an object, say a clock, and Sketchfab offers 50 different clocks, only three of which are suitable for an office setting.

To address this issue, we introduce the \textit{Object Retriever}, a skill that employs other AI models to identify the 3D object that the user most likely intended. The workflow of the \textit{Object Retriever} can be formalized as follows: given a user prompt $u$, the \textit{Object Retriever} identifies an object $o$ contained in $u$ and calls the Dall·E-2 \cite{dalle2} API for the object $o$, generating a "target image" $T$. Concurrently, the same object-prompt $o$ is used to download $N$ screenshots of 3D objects freely available on Sketchfab, denoted as $\bm{S} = \{s_1, s_2, ..., s_N\}$. We then employ CLIP \cite{radford2021learning} to map out similarity spaces in the language domain $L$ and the visual domain $V$. We select the top 5 images $\bm{S}' \subset \bm{S}$ that are closest to the object-prompt $o$ in the language similarity space $L$, and from these, we select the image $s^*$ that is closest to the target image $T$ in the visual similarity space $V$. Formally, let $L(o, s_i)$ and $V(T, s_i)$ denote the language and visual similarity between the object-prompt $o$ and the screenshot $s_i$, and the target image $T$ and the screenshot $s_i$, respectively. The \textit{Object Retriever} operates as follows:


This process is repeated to generate entire scenes. Algorithm \ref{3dobject_algo} and Figure \ref{fig:object-retrieve} describe this pipeline. There is potential for further exploration to improve this pipeline. For instance, selecting from the visual similarity space before the language similarity space might yield better results. Future work will involve human feedback to identify the workflow that maximizes the likeness between the 3D object loaded and the user's intended object.

\begin{algorithm}[hbt!]
\caption{Retrieving 3D objects}
\label{3dobject_algo}
\Input{$u$: user's prompt}
\Require{$o$: object in $u$; \\
$T$: target image; \\
$\bm{S}$: screenshots; \\
$L(o, s_i)$: language similarity; \\
$V(T, s_i)$: visual similarity.}

$\bm{S}' \gets \text{Top 5} s_i \in \bm{S} \text{with highest} L(o, s_i)$\;
$s^* \gets \arg\max_{s_i \in \bm{S'}} V(T, s_i)$\;
\Return $s^*$
\end{algorithm}

\begin{figure*}[hbt!]
    \centering
     \includegraphics[width=\textwidth]{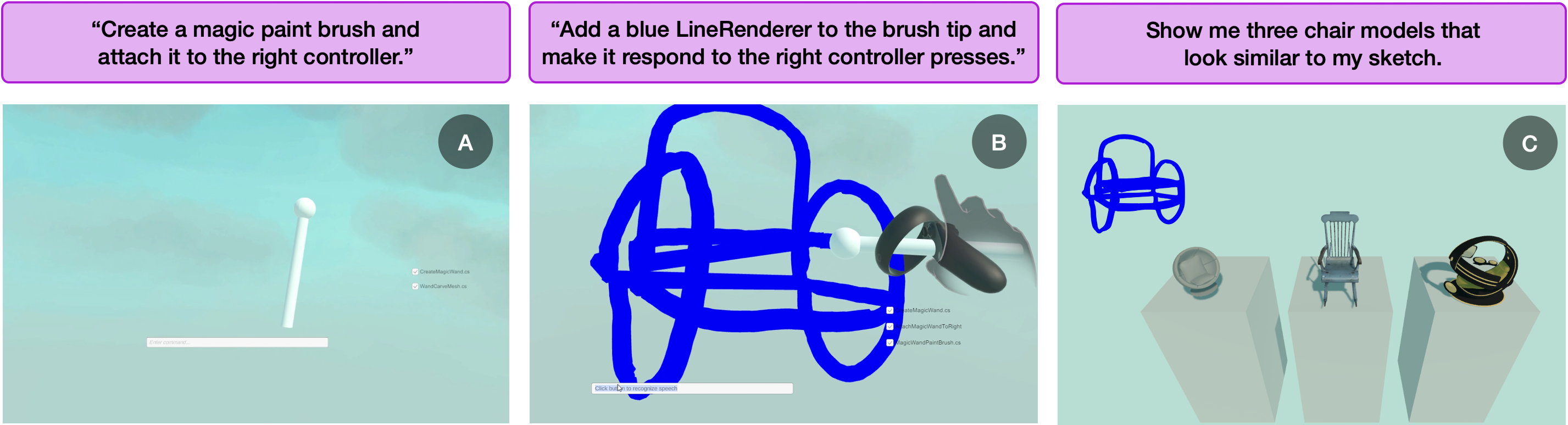}
    \caption{\textit{Sketching objects into existence with LLMR.} In the left panel, a user requests a "magic paintbrush" to be attached to a VR controller. The middle panel illustrates the automatic conversion of the line renderer into a paintbrush, where the user is shown drawing a chair. The right panel demonstrates the 2D-to-3D transformation using 2D-3D ControlNet \cite{zhang2023adding} and our Dall·E-CLIP Sketchfab API. This enables the generation of multiple chair models that can then be transferred across different platforms using LLMR for further interaction.}
    \Description{Sketching objects into existence with LLMR. Three-part image demonstrating the LLMR process for creating virtual objects. Panel A shows a virtual reality (VR) environment with a simple, white, floating paintbrush next to a VR controller, illustrating the user's request for a "magic paintbrush" to be attached to the controller. Panel B depicts the user's action within the VR environment, showcasing a complex, scribbled blue drawing in three-dimensional space, representing the automatic conversion of a line renderer into a functioning paintbrush. Panel C presents a variety of chair models in a 3D space, indicating the transformation from a 2D sketch to 3D objects using 2D-3D ControlNet and an API, which allows for the generation and interaction with multiple chair models in different VR platforms.}
    \label{fig:magicpaintbrush}
\end{figure*}

\section{Memory Management}
\label{sec:memory}

By default, language models generate new words based on all previously sampled tokens, a configuration that may not be ideal due to their finite context length. For instance, this may hinder the model's ability to engage in extended conversations. To mitigate this, techniques such as dialogue summarization and distillation can be employed \cite{dialogue_summarization,memory_management_long_term_convo, interactive_mem_management}. Additional research has delved into leveraging persistent memory and retrieving in-context examples from databases to enhance few-shot performance \cite{LLM_LTM, LLM_memory_bank}.

\begin{table}[b]
  \centering
    \begin{tabular}{|p{0.12\textwidth}|p{0.15\textwidth}|}
        \hline
     \textbf{Module} & \textbf{Memory Mode} \\  \hline 
     Planner & Memory-less\\   \hline
     Scene Analyzer & Memory-less\\   \hline
     Builder & Limited-memory\\   \hline
     Inspector & Memory-less\\   \hline
     Skill Library & Memory-less\\   \hline
    \end{tabular}
  \caption{Memory mode for each module. Note that no module uses full memory, the default GPT paradigm.}
  \label{tab:memory_mode}
  
\end{table}

We sought to deploy a protocol that alters the contents within the LLM's context window while the framework is in continuous use. We explored three memory modes for each module within \framework \ : full memory, limited memory, and memory-less. We document the memory modes used for each module in Table \ref{tab:memory_mode}. These modes pertain to the retention of all, a few, or none of the historical instructions and generated code within the model's context. Define an episode of interaction as the input and output to the module for a single user prompt to LLMR. To implement a memory-limited module, for example, we clear its context of all but the most recent $N$ episodes after every prompt, where $N=1$ typically. 


An effective memory management protocol offers three distinct advantages:

\textit{Token limit}: Trimming old memory reduces token consumption and enables prolonged usage of \framework, a critical feature for gradually constructing intricate scenes. Notably, the \textit{Scene Analyzer} benefits from having no memory of prior interactions, as it is susceptible to token constraints. As an example, the first AI2-THOR scene hierarchy measures around 7k GPT-4 tokens \cite{kolve2017ai2}. Hence, a full memory \textit{Scene Analyzer} with 8k tokens can only fulfill a single instruction before its context is depleted, rendering the framework essentially unusable outside of a memory-less setting.

\textit{Performance}: Certain modules perform better with reduced memory, as they may be prone to be confused by earlier interactions. For example, our empirical observations indicate that the \textit{Inspector} module exhibits increased leniency in repeated inspections, allowing the proposed code to pass before all errors are rectified.

\textit{Interpretability}: A memory-limited framework provides clearer error attribution. For instance, when a sequence of prompts is sent, and the generation fails at the final step, maintaining all memory makes it challenging to discern whether the last prompt posed a unique challenge or if the framework became perplexed by aspects of an earlier task. Improved transparency facilitates swift debugging and iterating on our framework.

We believe the choice of memory mode is a crucial aspect of any LLM orchestration pipeline, and our design choices may offer insights for the development of LLM systems beyond the task of creating virtual worlds.

\begin{figure*}[hbt!]
    \centering
     \includegraphics[width=\textwidth]{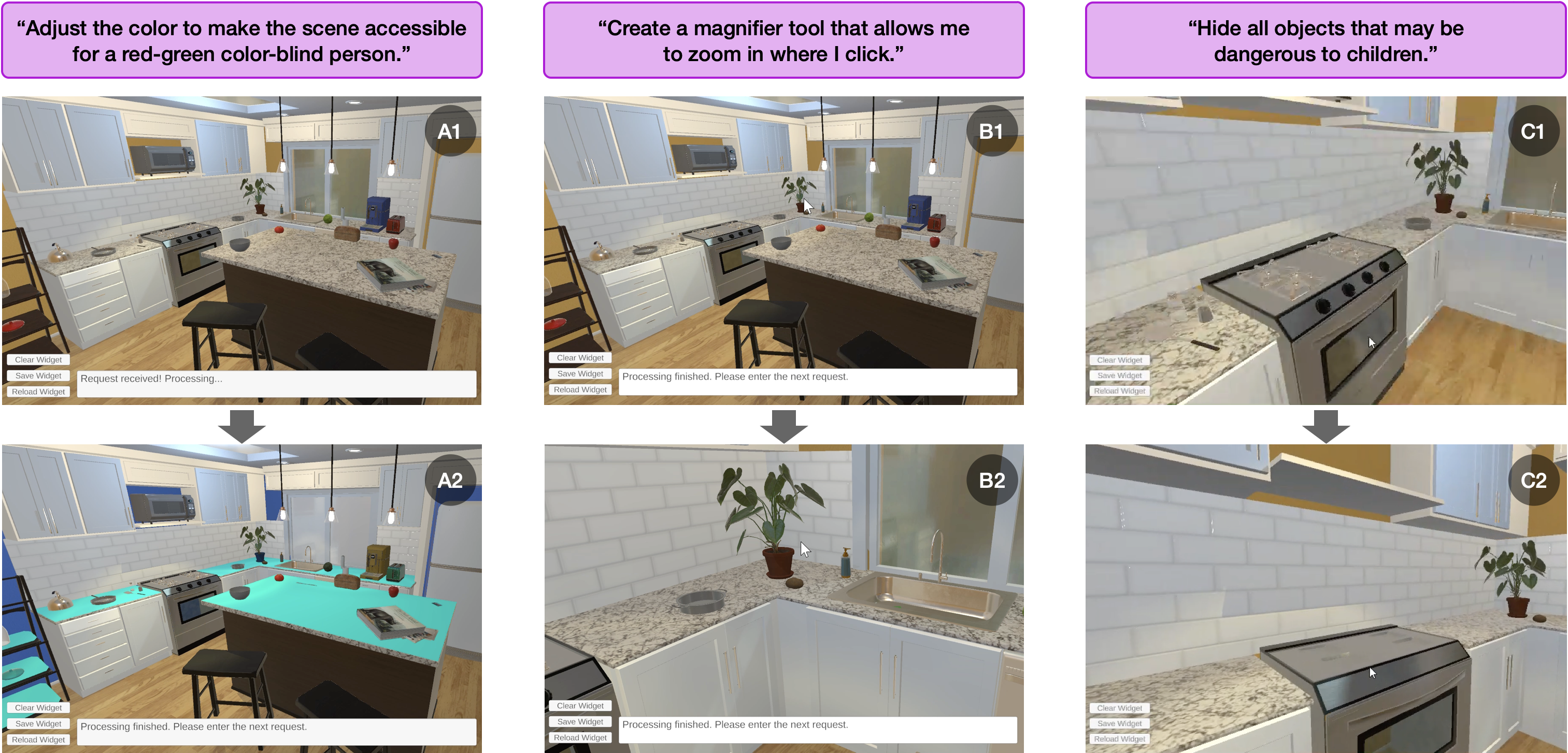}
    \caption{\textit{Accessible Interface Features in Action.} A1 and A2 show how a user can prompt the system to adjust the color scheme of a kitchen scene for red-green color-blind compatibility. B1 and B2 demonstrate the activation of a magnifier tool. C1 and C2 reveal the option to hide objects deemed not kid-friendly.}
    \Description{Accessible Interface Features in Action. Six-panel image sequence illustrating accessible interface features within a kitchen scene. Panels A1 and A2 display the same corner of a modern kitchen, with the color scheme adjusted from standard to a palette suitable for red-green color-blind individuals, showing a marked difference in contrast and color tones. Panels B1 and B2 highlight a digital magnifying tool in use; B1 shows a standard view of the kitchen counter and stove, while B2 zooms in on the stove controls for clearer visibility. Panels C1 and C2 focus on child safety features; C1 depicts the kitchen with various items on the counter, including knives, while C2 shows the same view with potentially dangerous objects like knives hidden from sight, creating a safer environment for children.}
    \label{fig:accessability}
    
\end{figure*}

\newpage
\section{Cross-Platform Compatibility and Installation}

We show that our framework can be deployed in various types of platforms (e.g., Web, Mobile, AR, and VR) and on various devices (e.g., Meta Quest, HoloLens 2). To keep the framework lightweight, we deploy our framework's run-time compiler on a PC that acts as the server, and we build upon existing remoting protocols and frameworks \cite{remoting, render_streaming} to stream the generated results to the client device (e.g., holographic remoting for a HoloLens 2). Platform dependencies, such as namespaces and other packages can be added as a "Skill" to the framework's \textit{Skill Library}, which allows the user to quickly enable interaction modalities such as pinch and input modalities like speech and controller.  

Interactive elements built within one scene can be saved as self-contained units by storing the source code that created them. We can then re-execute the cached code to load and adapt the prompted objects into novel scenarios, which can be as simple as a different scene with adjusted physics or a project with completely new APIs, as depicted in Figure \ref{fig:car-crossplatform}. Our experiments with \framework \ suggest that translating interactive elements between independent SDK platforms is possible and suggests an application of adapting existing pieces of software (perhaps ones written with obsolete, no-longer working code) to newer SDKs. We leave this for future explorations.


\subsection{Installation}

Our framework can be easily added to any existing Unity scenes. The framework consists of a unity package and a few additional open-sourced packages (such as GLTF loader and OpenAI), and the installation process takes only a few steps. This enables anyone with an OpenAI API key to try our framework. We are strong proponents of the adaptability of our framework, and so we have open-sourced the foundational framework along with several examples on GitHub (\url{https://llm4mr.github.io/}). Readers who wish to try our framework can try out the example playground scenes or can easily add our Unity package to their existing Unity projects. They would need to obtain an OpenAI API key, a copy of the Roslyn compiler and optionally an account for Sketchfab if they wish to automatically load existing assets. In the Appendix, we also provide the metaprompts used for each LLMR's modules for transparency.

\begin{figure*}[hbt!]
    \centering
     \includegraphics[width=\textwidth]{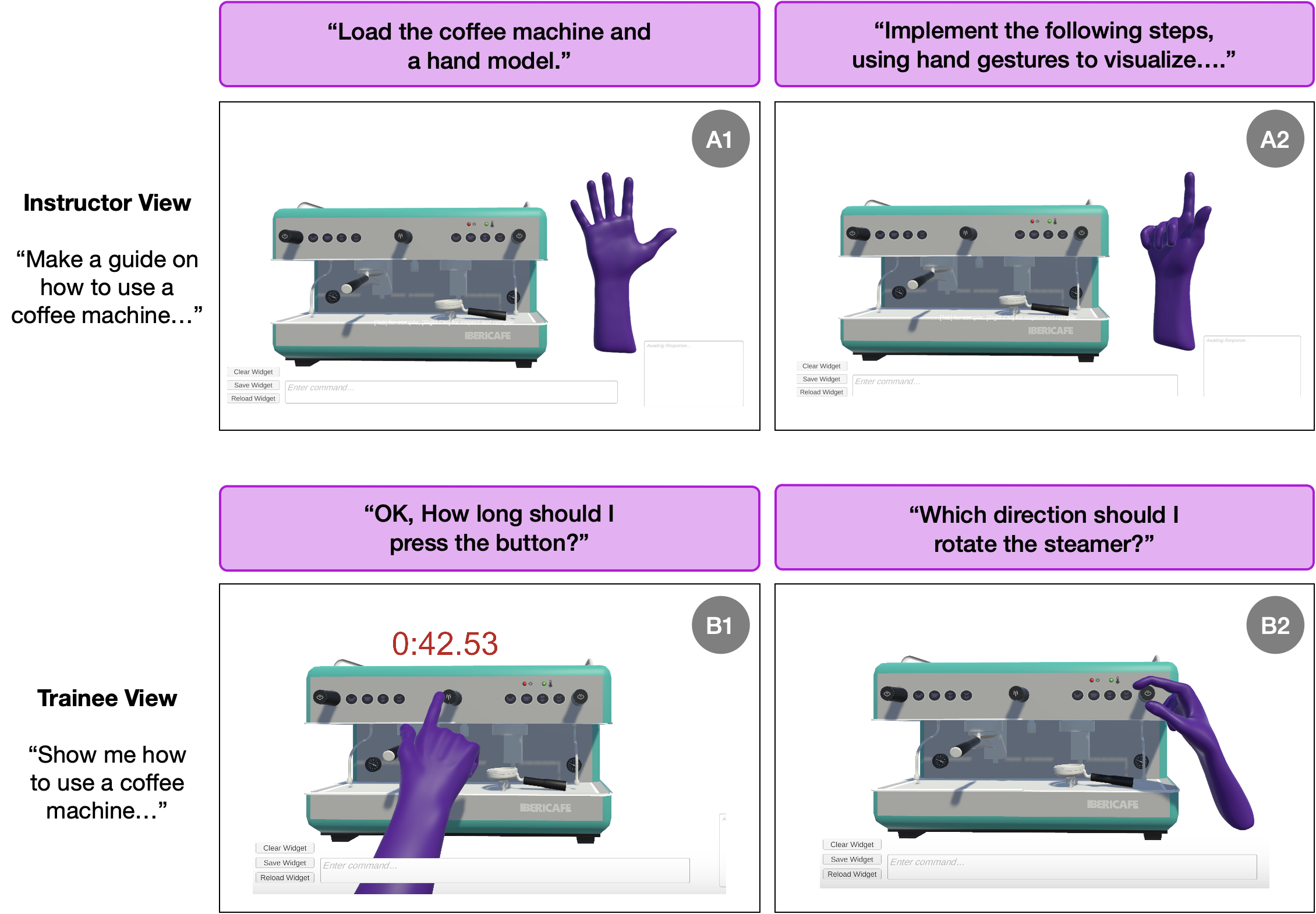}
    \caption{\textit{Spontaneous Creation of Teaching Guides}. A demonstration of creating a guide for operating a coffee machine in which LLMR animates a hand model to point out the various steps of the operation. Our framework allows for the rapid creation of such guides and furthermore allows users to ask questions that were not predicted by the instructor beforehand, with appropriate motions being animated on the fly.}
    \Description{Spontaneous Creation of Teaching Guides. Four-panel image sequence demonstrating an interactive guide for operating a coffee machine with LLMR technology. The top two panels, labeled A1 and A2, represent the instructor's view. A1 shows a coffee machine interface with a virtual hand model hovering over a button, indicating the first step with the text prompt "Load the coffee machine and a hand model." A2 depicts the hand model gesturing to a steamer knob with the text "Implement the following steps, using hand gestures to visualize..." The bottom two panels, labeled B1 and B2, depict the trainee's view with a timer and query responses. B1 shows the hand model pressing a button with a timer displaying 42:53 and the caption "OK, How long should I press the button?" B2 illustrates the hand model turning the steamer knob with the question "Which direction should I rotate the steamer?" This visual aid conveys the dynamic nature of the teaching guide, with a system that animates hand motions in response to user queries, facilitating on-the-fly instructional interactions.}
    \label{fig:teachingguide}
    
\end{figure*}

\section{Example Prompted Interactive Worlds and Uses}

In this section, we illustrate the wide range of objects, tools, and scenes one can construct with \framework. We highlight that our framework is modular, real-time, adaptive, interactive, and multi-modal, which differentiates this approach from other generated 3D worlds that primarily focus on visual appearance. For all of the examples below, it is important to stress that all of the results are achieved simply by prompting the system, without the need for manual intervention.

\newpage
\subsection{Game Design and Creativity}

An immediate application of our framework is the creation of games, in particular, scenes. A scene sets the context of a game, and it usually involves numerous assets that are difficult and tedious to set up manually. A game designer can use the \textit{Planner} to create a draft environment, and add interactive components like "players" and "opponents" with responsive behaviors to mock up the gameplay logic. 
In addition, game designers can expand gameplay in multiple environments. For example, a toy car can be created and reloaded in a moon simulation environment in VR (Figure \ref{fig:car-crossplatform} B) or be spawned in the physical world and driven around with a mobile phone (Figure \ref{fig:car-crossplatform} C). Besides "prompting" objects into existence, we show that our framework also allows users to "draw" things into existence. Here the user wishes to design a chair (Figure \ref{fig:magicpaintbrush}).  They can do so by simply prompting "a magic paintbrush", which has functions similar to that of TiltBrush \cite{tiltbrush}, a popular 3D drawing application, and then turn the drawing into a 3D model with the integration of Dall-E 2, CLIP, and Sketchfab, through a similar process illustrated in Figure \ref{fig:object-retrieve}.

\begin{figure*}[hbt!]
    \centering
     \includegraphics[width=\textwidth]{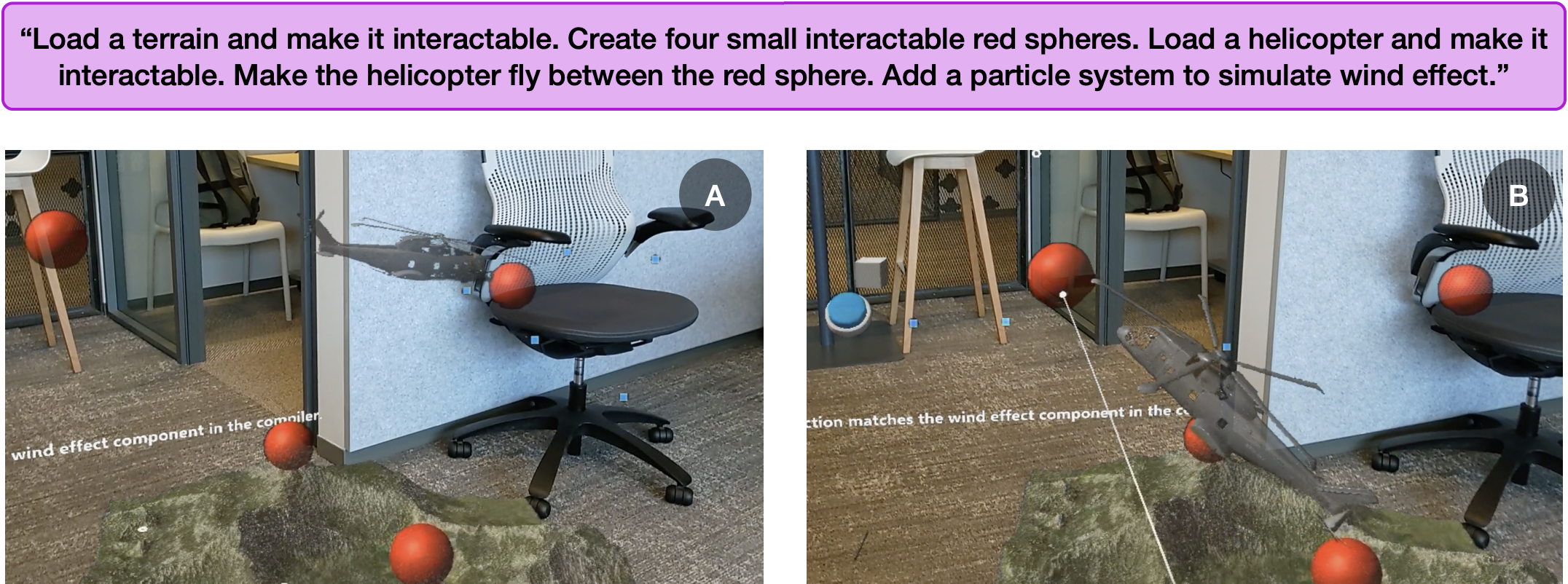}
    \caption{\textit{Simulated Rescue Plan.} The HoloLens 2 displays the automated generation of a simulation of a rescue plan using our framework. The guide shows an interactable 3D terrain, helicopter, and simulated wind, allowing rescue workers to visualize the flight path under different weather conditions.}
    \Description{Simulated Rescue Plan. The image shows a side-by-side comparison labeled as (A) and (B). Both depict a simulated rescue plan using augmented reality on a HoloLens device. Description for (A): A virtual 3D environment is overlaid on a physical office setting. Four small, red spheres are arranged in a square pattern on the floor. A virtual helicopter is positioned near the center, slightly above a black office chair. The text instruction suggests interaction with the terrain and helicopter, and the presence of a wind effect represented by a red arrow near the bottom left sphere. Description for (B): Similar to (A), with the virtual helicopter now located closer to the rightmost red sphere. The red arrow indicating the wind effect is more pronounced and closer to the sphere, which could suggest the helicopter has moved due to the simulated wind.}
    \label{fig:remotecontrol}
\end{figure*}

\subsection{Accessibility and Adaptive Interface}

Similar to the accessibility feature in 2D documents, our framework can also be prompted to make a 3D scene accessible and adaptive to different user needs and preferences. Figure \ref{fig:accessability} shows three examples of editing an existing virtual kitchen scene to different requests. For example, one can request to make the scene to be more friendly to red-green color-blind users. For someone who is near-sighted, they can prompt a magnifier tool that zooms into a particular part of the room. An architect can use our framework to figure out if the space is friendly for wheelchair users or make sure objects in the room are child-proof. These examples show how our framework leverages LLMs' prior knowledge and puts the knowledge into the context of a spatial world at a human scale.

\newpage
\subsection{Remote Assistance and Planning}
In a remote training scenario, typically, creating such a 3D interactive training guide requires custom creation, from rigging a gesture to placing a UI element. An instructor can use our framework to automate the generation of a training guide from a list of instructions.  (Figure \ref{fig:teachingguide}).  The trainee can then, for example, use an AR device that overlays information on the machine. As the trainee advances through the steps, they can ask questions directly to the guide where answers can be generated in the context of the trainee's learning progress. In another scenario of remote rescue planning, helicopter operators can prompt a simulation of the flight path given several target locations and see how the flight path might be affected by different wind conditions (Figure \ref{fig:remotecontrol}).

\begin{figure*}[hbt!]
\centering
     \includegraphics[width=\textwidth]{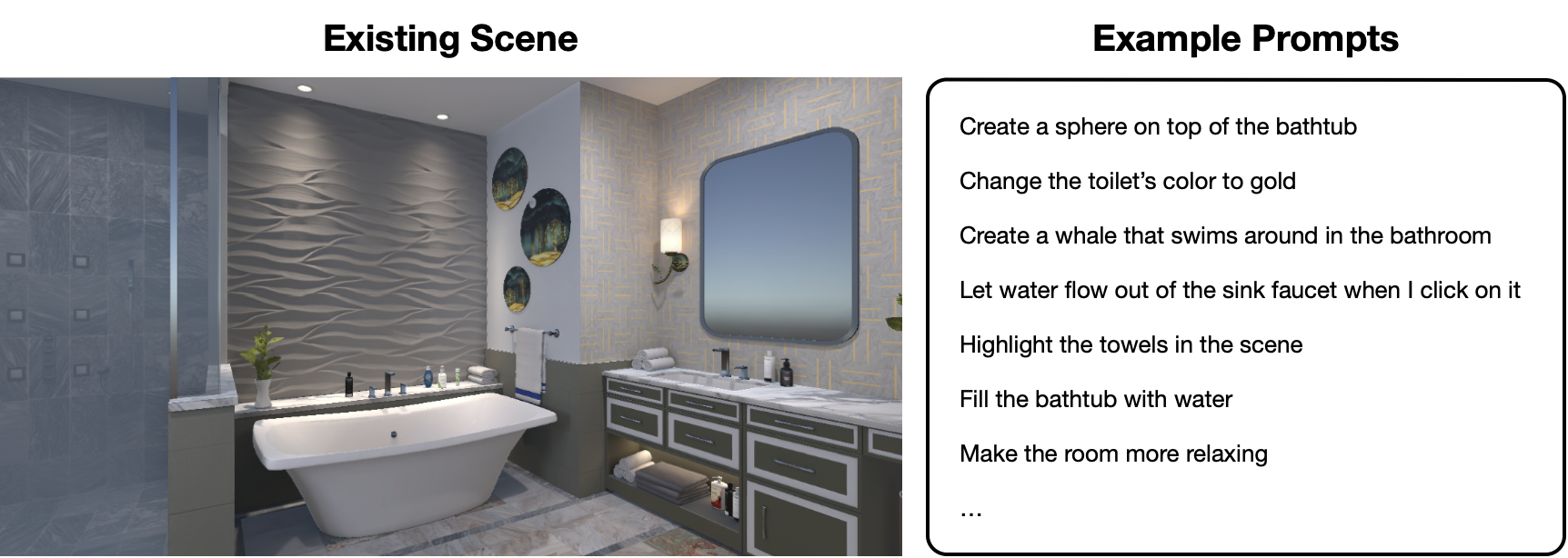}
\caption{\textit{An illustration of our experimental setup.} We provide the bathroom scene (left) and a subset of the 150 prompts (right) used in this space for the evaluation provided in Figure \ref{fig:numerical_study_error_rate}. }
\Description{An illustration of our ablation experimental setup. The image is split into two sections: "Existing Scene" on the left and "Example Prompts" on the right. The photo displays a modern bathroom scene with a neutral color palette. A freestanding white bathtub is placed in the center, with a grey-tiled wall behind it featuring three recessed shelves adorned with plants and towels. To the left of the bathtub, a towel radiator is mounted on the wall, and to the right, a large mirror hangs above a sink with visible faucets. The floor is also grey, complementing the wall tiles. On the right side, a list of example interactive prompts is displayed against a white background. The prompts include actions such as creating a sphere on top of the bathtub, changing the toilet's color to gold, and letting water flow out of the sink faucet when clicked. Other prompts suggest more complex interactions like filling the bathtub with water and making the room more relaxing, indicating a dynamic and interactive virtual environment.}
\label{fig:Bathroom_scene}
\end{figure*}

\section{Numerical Study}

As an orchestrated pipeline, LLMR augments an LLM coder with multiple modules to enhance its reliability. To empirically justify the inclusion of each module in our framework, we quantitatively evaluate LLMR's generative performance against a variety of prompts and baselines. In addition to success in compiling the generated code, we evaluate how our framework meets our design goals: real-timeness, complexity of interaction, and iterative fine-tuning ability. 

This section is organized as follows: we begin by evaluating LLMR on single prompts in an empty and existing scene, highlighting the impact of each module and overall performance compared to standard LLMs. We also discuss the framework's performance at completing tasks with different complexity. Then, we conduct a similar experiment on sequential prompts to illustrate LLMR's capacity for iterative designs. Lastly, we present an analysis of the real-time aspect of our framework.

\subsection{Error Rate}

\subsubsection{Experiment Setup}
We start by investigating LLMR's ability to carry out single, independent requests in either an empty or existing scene. To this end, we created two datasets each with 150 prompts. The first set is used as inputs in an empty scene and is mainly creative in nature as there is nothing to modify or interact with in the world. An example is "creating a cat and mouse out of primitives. The cat should chase the mouse, who flees in an erratic pattern." The second set is used as inputs in an \textit{existing} scene shown in Figure \ref{fig:Bathroom_scene}. The scene was downloaded from Sketchfab \cite{sketchfab} and was chosen as it is sufficiently complex (around 35 objects). A few example prompts are shown in Figure~\ref{fig:Bathroom_scene}, which involve visual and semantic alteration of the space. To promote fairness and diversity in our test prompts, we use a separate, properly prompted GPT to generate two evaluation datasets. The authors created 15 prompts as demonstrations for the prompting GPT. The full evaluation datasets can be found in \textbf{Appendix}.

To assess the efficacy of LLMR, a proper metric is required. Given the subjective nature of tasks such as "make the room more uplifting," it is difficult to systematically determine if a prompt has been met successfully. However, the presence of run-time or compilation errors in the generated code can be considered a clear indicator of failure. Therefore, we have selected the 'error rate' – the proportion of outputs with bugs – as the criterion for assessing the framework's performance.

To evaluate the efficacy of each module of LLMR, we created three model conditions, each with adding one additional GPT module, besides GPT-4 zero shot and GPT-4 few shot as our baselines. This makes a total of 5 model conditions. We conducted 5 runs of 150 prompts with each model condition and for each scene condition (empty scene and existing scene).

\newpage
\subsubsection{Results and Discussion}

\begin{figure*}
    \centering
     \includegraphics[width=\textwidth]{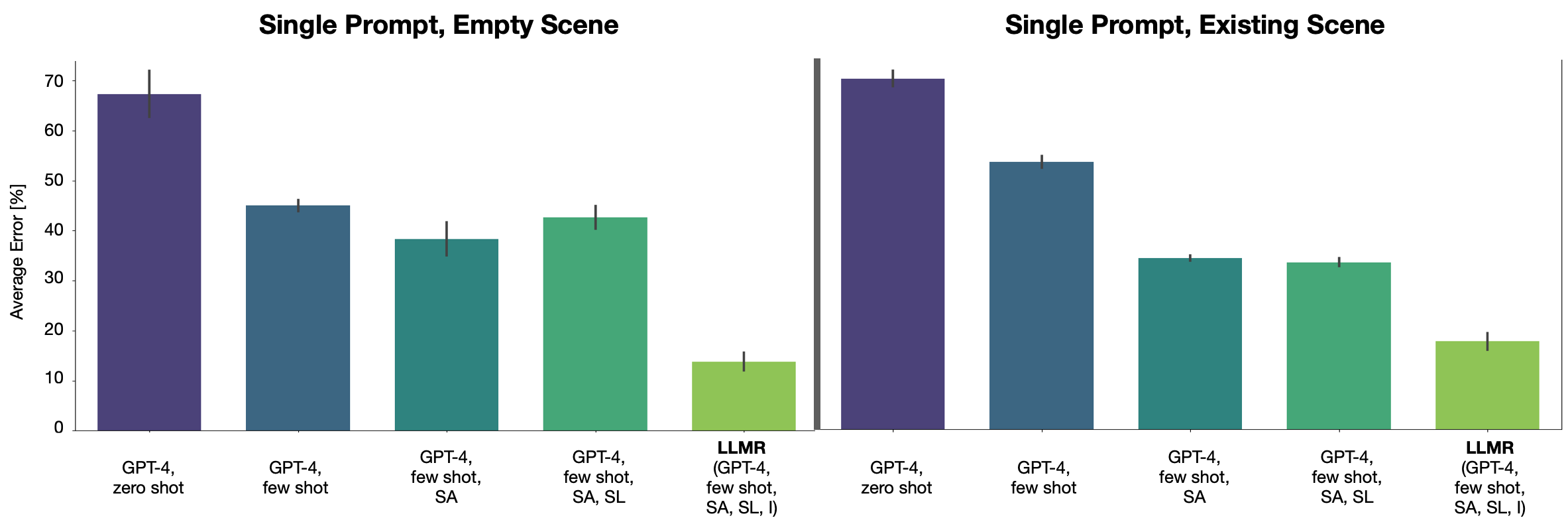}
    \caption{\textit{Comparison for average compilation and run-time error rate.} SA stands for the \textit{Scene Analyzer}, SL stands for the \textit{Skill Library}, and I stands for the \textit{Inspector}. Overall, in both creating from scratch, as well as editing existing scenes, \framework \ outperforms GPT-4 by 3x in the case with few-shot prompting, and gives over 4x improvement compared to the performance of zero-shot GPT-4.}
    \Description{Comparison for average compilation and run-time error rate. The image presents two bar charts side by side, comparing the average compilation and runtime error rates across different methods. Description of the first bar chart, labeled "Single Prompt, Empty Scene": The chart displays five bars, each representing different combinations of techniques and prompting methods with GPT-4. The bars are shades of grey and orange, with the orange bars representing the LLMR (GPT-4, few-shot, SA, SL) method. Each bar is annotated with a mean (M) value at the top and a standard error (SE) value just below it. Description of the second bar chart, labeled "Single Prompt, Existing Scene": Similar to the first chart, it has six bars with the same color scheme and annotations.}
    \label{fig:numerical_study_error_rate}
\end{figure*}

We provide a summary of error rates for our model and various baselines in Figure \ref{fig:numerical_study_error_rate}. To underscore the benefit of each LLMR module, we add each component incrementally to tease out its marginal impact. Starting with the off-the-shelf GPT-4, we see that standard in-context learning techniques increase performance in both settings, yet only to the extent that roughly half of the requests fail. From here, we augment the standard GPT-4 with components developed in this work, starting with the \textit{Scene Analyzer}, then the \textit{Skill Library}, and finally the \textit{Inspector}. As a result, the generated errors drop substantially to only $20.5\%$ and $25.2\%$ of the error rate observed in the original GPT-4 for the empty and existing scene, respectively, which attests to the effectiveness of our pipeline over standard, off-the-shelf LLMs for the task of generating interactive scenes.

We now discuss the impact of each LLMR module in detail. As explained in Section \ref{sec:SA},the \textit{Scene Analyzer} allows LLMR to parse and understand the virtual scene and is thus indispensable for meaningful manipulations of existing environments. Consequently, enhancing GPT-4 with the \textit{Scene Analyzer} results in a significant performance enhancement in the Bathroom scene. Secondly, the \textit{Inspector} module enables LLMR to perform self-debugging and effectively prevents the generation of erroneous code, further reducing the error rate in both scenarios. Although we integrated the \textit{Inspector} at the final stage, it is compatible with any combination of modules and will consistently reduce the output error rate. As an example, we added \textit{Inspector} to GPT-4 with few-shot prompting in the empty scene and observed the average error drops from 45.0\% to 13.1\%. We also observe the \textit{Skill Library} has a marginal impact on the error rates. This is expected, since the \textit{Skill Library} is designed to handle more specialized tasks, which we discuss in more detail in the following subsections. Lastly, the \textit{Planner} is not included as it alters the input prompt with step-by-step decomposition, making the results incommensurable. We include in the Appendix an example where the \textit{Planner} is used to build a virtual kitchen, underscoring the benefit of decomposing difficult tasks into incremental steps.

\subsection[Error Rate by Levels of Difficulty]{Error Rate by Levels of Difficulty}


\subsubsection{Methodology}
To explore the relationship between the complexity of prompts and the completion rate of LLMR, we performed an ad-hoc analysis of the results from the previous section. We classified our prompts for single prompt task into levels of difficulty from 1 to 10. To achieve this, we utilized a non-contextualized LLM devoid of any meta-prompting, asking it to assign a difficulty level to each prompt. This process was repeated ten times for each prompt, and the average difficulty level was then calculated (one which had a small standard deviation). The aggregated results, categorized by difficulty level, are illustrated in Figure \ref{fig:levels_ablation}. The prompt given to this LLM (GPT-4) was "\textit{The above are prompts that are given to a system that can code and execute commands inside of Unity. We want to measure how good this system is at coding in C\# for Unity purposes. Given your knowledge of Unity, please rate all of the prompts above on a level of difficulty from 1 to 10}". The rationale behind employing a non-contextualized LLM (without any meta-prompting) lies in the subjective nature of assessing difficulty levels. Being the developers of the system, our judgment might be inherently biased, influenced by our understanding of the system's capabilities and limitations. Furthermore, engaging Unity experts to determine the difficulty levels presents its challenges. The variability in the expertise and experience levels among Unity developers could lead to inconsistent evaluations and difficulty in standardizing the experience of the evaluators without a comprehensive and uniform examination framework.

\subsubsection{Results and Discussion}

\begin{figure*}[hbt!]
    \centering
     \includegraphics[width=\textwidth]{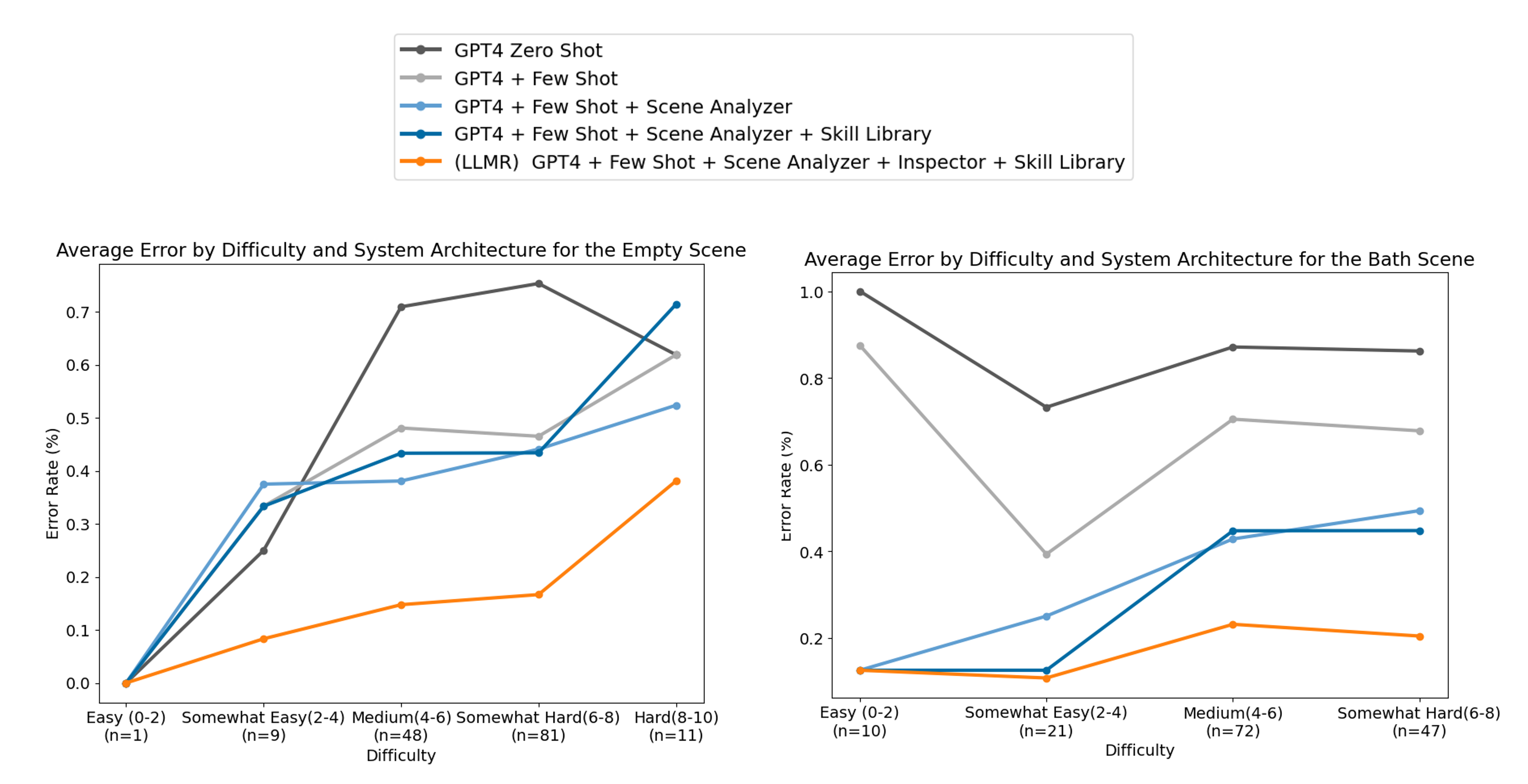}
    \caption{\textit{Performance Improvement of LLMR Modules Organized by Difficulty Level of Prompts}. Comparing Error Rates of GPT-4 with Incremental LLMR Module Integration. The error rates for most methods increase with difficulty, and the LLMR method (in orange) still maintains a consistently lower error rate compared to others.}
    \Description{Comparison for average compilation and run-time error rate. The image contains two line graphs placed side by side. Description of the first graph, labeled "Empty Scene": This graph plots the average error rate on the y-axis, which ranges from 0 to 0.7, against varying levels of difficulty on the x-axis, labeled from "Easy (1-2)" to "Hard(10)". There are four lines representing different methods: GPT-4 Zero Shot, GPT-4 + Few Shot, GPT-4 + Few Shot + Scene Analyzer, and LLMR (GPT-4 + Few Shot + Scene Analyzer + Inspector + Skill Library), each marked in different colors and with corresponding symbols. The LLMR line (in orange) demonstrates a generally lower error rate across all difficulty levels. Description of the second graph, labeled "Existing Bathroom Scene": This graph has a similar structure, with the average error rate on the y-axis and difficulty on the x-axis. It shows the performance of the same four methods. In this scenario, the error rates for all methods increase with difficulty, but the LLMR method (in orange) still maintains a lower error rate compared to others, although the differences are less pronounced compared to the "Empty Scene" graph.}
    \label{fig:levels_ablation}
\end{figure*}

In this section, we analyze the performance of various architectures in executing Unity tasks, differentiated by difficulty levels that range from Easy to Hard. These levels were determined based on a 1-10 scale assigned by GPT-4. Figure \ref{fig:levels_ablation} shows the error rate of the different architectures on two panels. On the left, we have the results for the empty scene and on the right for a scene with a bathroom containing various objects.

Across all levels and scenes, LLMR (orange line) consistently outperforms other architectures, underscoring its robustness. In the empty scene on the left, a noticeable trend is that the error rate generally increases with the task difficulty. This trend aligns with expectations, except in the case of GPT-4 Zero Shot. A notable point here is that the Easy category only contains a single prompt, which is a basic "Hello World" console display. The simplicity of this task explains its solitary placement in this category. For the bathroom scene, the error rates for Medium and Somewhat Hard tasks show minimal variation, suggesting a plateau in difficulty perception. An interesting observation is the drop in error rates from Easy to Somewhat Easy tasks, although this is not consistent across all models. The integration of the \textit{Skill Library} shows mixed effects (dark blue line). In some instances, it enhances performance, while in others it seems to hinder it.

Estimating the difficulty of tasks, especially in scenarios involving modifications to an existing scene rather than building from scratch, presents challenges. This is exemplified in the bathroom scene, where adding new objects (difficulty levels 3-4) did not require scene understanding, contrary to the tasks in the Easy category, which involved moving objects and thus relied more on scene comprehension. Our analysis of the prompts indicates that the nature of the scene significantly influences the perceived difficulty. For instance, in the bathroom scene, certain tasks categorized as Easy in theory turned out to be more challenging in practice. The \textbf{Appendix} offers a more comprehensive analysis, including variations in architecture, such as the combination of \textit{Scene Analyzer} (SA) and \textit{Inspector} modules.

In conclusion, LLMR demonstrates superior performance across various scenarios, underscoring its effectiveness in handling tasks of varying complexity in Unity environments. This analysis also highlights the intricate relationship between task difficulty, scene context, and architectural components, paving the way for further exploration in optimizing task-specific architectures.


\subsection[Task Complexity]{Task Complexity}

Complexity can manifest through different aspects. To supplement the ad-hoc analysis above, we now provide a more comprehensive discussion by breaking down the concept of complexity through the following aspects and share findings that emerged throughout our experimentation:

\textit{Specific Skill Requirement} – Certain tasks are inherently more difficult. For example, deforming the mesh of an object is much more complicated than adding an object to the scene. A human developer may need to look up examples and documentation to achieve a complex task; LLMR can reduce the complexity of the task by starting a templated script. However, LLMR is not error-proof. As shown in the previous section, LLMR's error rate increases (but not above 40\%) as a task becomes more difficult. The error rate can be further brought down by adding relevant skills to the \textit{Skill Library} by an experienced developer to help LLMR achieve a higher success rate and reduce possible rounds of iteration between the \textit{Builder} and \textit{Inspector}. LLMR can save the time of experienced developers by generalizing beyond the examples provided in the \textit{Skill Library}.

\textit{Token Requirement} – The amount of tokens required grows as the scene or the object becomes more complex. For example, if the existing scene has a tree object with many leaves, where each leaf is considered a child game object, the scene summary could easily exceed the maximum token allowed (at the time of writing, the maximum number of tokens was 8k, though this has now grown significantly). In anticipation of this, our \textit{Scene Analyzer} module only fetches the top-level game object name to filter out the relevant game object based on user query and task. This allows our framework to handle a scene as complex as the Kitchen scene (example in Figure \ref{fig:accessability}) and the Bathroom scene (used in Numerical Study, Figure \ref{fig:Bathroom_scene}). Besides the complexity of objects and scenes, a task itself can also require a lot of tokens. One such example is generating an animation \cite{huang2023real} (with the help of a skill written for the \textit{Skill Library}) that involves generating time-series of numerous joint positions and rotations. An optimization could involve developing a simple way to represent the time-series information.

\textit{Memory Requirement} – When a task requires prior knowledge of previous prompts (e.g., behaviors created by previous prompts), this requires previous prompts to have been successfully compiled and to be robust enough. We described approaches to managing the memory of the different modules in Section 5 to both conserve total memory consumed while preserving the necessary information for the framework to carry out complex tasks.

\textit{Quality Requirement} – A user may request different levels of fidelity of the output. For example, the user could create a complex scene out of primitives only with the help of the \textit{Planner} module (e.g., a full kitchen, Figure \ref{fig:teaser}) instead of out of higher fidelity 3D models (see examples of participants' creations in the video figure). The flexibility to create visually simple yet functional and interactive scenes is akin to creating a lo-fi mockup that allows users to quickly prototype and iterate without waiting for the full generation of 3D scenes that are visually complex but cost a lot of compute and time and are not easy to modify.

\begin{table*}[hbt!]
  \centering
  \begin{tabular}{|p{0.25\textwidth}|c|c|c|}
    \hline
     Model & Error rate ($\downarrow$) & Avg. prompt completion ($\uparrow$) & \% of fulfilled prompts ($\uparrow$)\\
     \hline
    GPT-4 & 0.745 & 0.339 & 0.288  \\
    \hline
    GPT-4, few-shot & 0.452 & 0.624 & 0.500 \\
    \hline
    GPT-4, few-shot with SA & 0.374 & 0.691 & 0.575  \\
    \hline
    \framework & \textbf{ 0.245} & \textbf{0.824} & \textbf{0.775}\\
    \hline
  \end{tabular}
  \caption{Numerical results for sequential prompts. The arrows next to the metrics point to its favored direction. For example, the down arrow next to error rate means a lower error rate should be preferred. }
  \label{tab:sequential_results}
  \vspace{5px}
\end{table*}

\begin{table*}[hbt!]
  \centering
  \begin{tabular}{|p{0.25\textwidth}|p{0.13\textwidth}|p{0.13\textwidth} |p{0.13\textwidth}|}
    \hline
     Model & Single prompts, empty [sec] & Single prompts, bathroom [sec] & Sequential prompts [sec] \\
    \hline
    GPT-4 & 35.24 & 20.60 & 77.10\\
    \hline
    GPT-4, few-shot & 37.82 & 21.28 & 112.50 \\
    \hline
    GPT-4, few-shot with SA & 33.90 & 20.58 & 69.40 \\
    \hline
    GPT-4, few-shot with SA, SL  & 34.46 & 21.64 & 74.60 \\
    \hline
    \framework & \textbf{90.98} & \textbf{49.16} & \textbf{170.90}  \\
    \hline
  \end{tabular}
  \caption{Average time taken in seconds to generate and compile each prompt. SA stands for the \textit{Scene Analyzer}, and SL stands for the \textit{Skill Library}. LLMR is equivalent to GPT-4 augmented with the \textit{Analyzer}, the \textit{Skill Library}, and the \textit{Inspector}.}
  \label{tab:realtime}
\end{table*}

\subsection[Iterative, Incremental Design]{Iterative, Incremental Design}

\subsubsection{Experimental Setup}
In practice, creating content-rich virtual worlds requires incremental steps. Therefore, it is important to assess how LLMR performs in iterative scenarios, where requests are made and fulfilled one after another to gradually build and alter a virtual scene. We tested LLMR with 80 sequential prompts, each averaging 5 single prompts. These sequential prompts consist of a set of instructions aimed at completing a complex task. For instance, a sequential prompt for constructing a bedroom might include steps like \textit{"create an empty room with walls; add a bed with a lamp next to it; add a window on the wall.}"

We use three metrics to evaluate performance in an iterative setting. First, the error rate on all individual prompts is considered and is the same as in single tasks. Second, we calculate the average degree of completion, measured as the number of completed single prompts over the sequence length for each sequential prompt. As the sequential prompts have varying lengths, accessing the completion average prevents "long and simple" sequences from flooding the error rate. Lastly, we define \textit{fulfilled prompts} to be sequential prompts that are completed from start to finish and compute their percentage over the total number of prompts. This is a demanding metric that validates whether the model can manage extended use sessions gracefully. In extreme cases, a model excelling only in short sequences can have a reasonable error rate yet zero perfectly fulfilled prompts.

\subsubsection{Results and Discussion}

The numerical results, presented in Table~\ref{tab:sequential_results}, show that GPT-4's performance in sequential tasks improves significantly with the addition of each LLMR module. When all modules are integrated, LLMR surpasses the standard GPT-4 by approximately 2.5 times across all metrics, aligning with results from single prompt tests. Furthermore, LLMR's memory-efficient design maintains a constant context usage for arbitrary prompt sequences and thus removes token size limitations during prolonged sessions. As such, LLMR demonstrates promising performance in the progressive creation and modification of virtual scenes, a scenario that resonates more closely with practical use cases. Lastly, we discuss in section \ref{sec:user_study} how the users subjectively rate the iteration process working with our framework.

In general, sequential prompts are much more challenging than single prompts because they require the model to maintain and manage long-range dependencies, a task known to be challenging in sequence modeling~\cite{rnn_long_range_dep}. To use the provided example, adding a window on the wall requires knowledge of the wall that was created a few prompts prior. From this perspective, the \textit{Scene Analyzer} serves as an effective summarization~\cite{dialogue_summarization} that helps the model redirect its attention to the part of the scene most relevant to the request, thereby reducing potential errors. In addition, the \textit{Inspector} receives scene parsing from the \textit{Scene Analyzer} and can thus effectively shield the generated code against potential errors in a sequential setting.

\subsection[Real-time]{Real-time}


\subsubsection{Methodology}

Last but not least, an important strength and design goal of our framework is the \textit{real-time} creation and modification of objects and scenes, which is crucial to ensure the practicality of use. To evaluate the framework's real-timeness, we measured the time taken for task completion (from generation to compilation) with different combinations of modules. Once again, the results were from 5 runs of 150 prompts with each model condition. Note that we ran the experiment from August 2023 until December 2023, where the performance and latency of OpenAI's GPT-4 model varied slightly but the difference was marginal. The experiments were run on a PC with 32GB of RAM and an Nvidia RTX 3080 GPU.


\newpage
\subsubsection{Results and Discussion}
Table \ref{tab:realtime} shows the average completion time (i.e., including generation and compilation time) for each model and condition. The off-the-shelf GPT-4 takes around half a minute to complete a single prompt, and the full LLMR framework on average takes a little over a minute – a timeframe we consider acceptable given the task complexity (Figure \ref{fig:levels_ablation}) and improved task completion rate (Figure \ref{fig:numerical_study_error_rate} and Table \ref{tab:sequential_results}). To put things into context, completing these complex tasks manually takes much longer even for someone reasonably familiar with Unity if we account for the time spent on looking up documentation and debugging. 

There are a couple of factors that contribute to the additional operation time. First, the complexity of certain tasks, such as retrieving 3D assets from Sketchfab, requires extra time to download assets from third-party sites. The time needed to finish retrieving a 3D model varies a lot and depends on the size of the model, and thus we did not include this in our evaluation. In this case, our framework anticipates this by caching the previously saved model for faster reloading. Second, to ensure the success rate of our framework, the Inspector nearly doubles the code generation time (see the last two rows of Table \ref{tab:realtime}). This is an inherent tradeoff, and the \textit{Inspector} module can be turned off for simple tasks. Finally, back-and-forth interactions between LLMR and the user as well as iteration over the generated results contribute to the overall development time. It is worth noting that during our user study (Section 9), none of the participants mentioned or complained about generation time. Participants who are novice Unity users appreciated that LLMR saved them time from the steep learning curve. In addition, our "saving and reloading" capability (Section \ref{sec:save_reload}) allows users to iterate faster by reusing prior creations, which takes less than 10 seconds to recompile.


\begin{figure*}[hbt!]
    \centering
     \includegraphics[width=\textwidth]{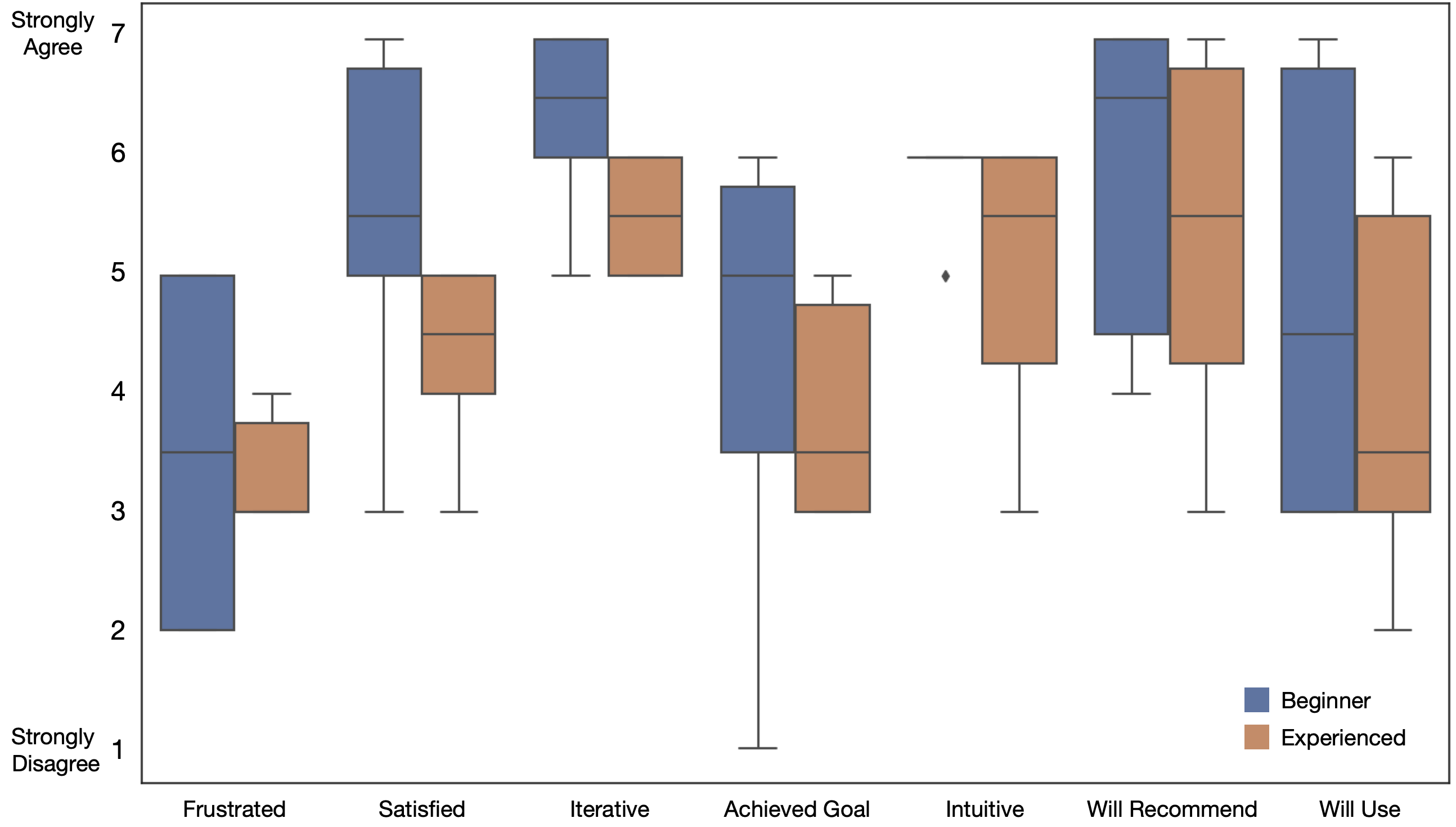}
    \caption{Results of the user study for both experienced and beginner users of Unity. Overall, the users found \framework \ satisfactory and would recommend others to use it too.}
    \Description{Results from user study. The image depicts a bar chart evaluating user satisfaction with Unity software, differentiated between beginner and experienced users across various criteria. Description: The chart has a 1 to 7 scale, where 1 stands for 'Strongly Disagree' and 7 for 'Strongly Agree'. There are seven pairs of bars, each representing a different aspect of user experience such as 'Frustrated', 'Satisfied', 'Iterative', 'Achieved Goal', 'Intuitive', 'Will Recommend', and 'Will Use'. Each aspect is evaluated by two bars, one for beginner users (lighter shade) and one for experienced users (darker shade). The bars range in height, reflecting the mean score given by the users, with error bars indicating the standard error. For example, on 'Satisfied', beginners scored around 5.5 and experienced users scored slightly lower, near 5.3. The 'Will Recommend' aspect has one of the highest scores for both groups.}
    \label{fig:user-study}
\end{figure*}

\newpage
\section{Usability Study}\label{sec:user_study}

The ablation test focused on only the compilation and run-time errors in the code generated by our framework. We also wanted to evaluate the quality of the generated output with human users. In addition, we also wanted to understand how users with different levels of familiarity with Unity would use our framework.

\subsection{Procedure}

We recruited twelve users (1 pilot, 11 participants) with different levels of experience using Unity (5 participants had more than one year of Unity experience). The participants' backgrounds were software engineers, product managers, or researchers. Each session took around 2 hours, and each participant had at least 1.5 hours to experiment with the framework. We provided a unity package that includes basic features (\textit{Scene Analyzer} and \textit{Skill Library,} \textit{Builder}, and \textit{Inspector}). Before the study, each participant downloaded the package to an empty or existing Unity scene and followed the instructions to set it up. Each participant went through a few rounds of interaction with the framework. A round of interaction could look like the following. The participant types: “Create a tool that changes the color of the car.” The framework processed the prompt and generated scripts that were then automatically compiled at run-time. The participant looked at the generated output and decided on the next prompt. The investigator might suggest different things to try or remind the participant of the capabilities of the framework. They were asked to think out loud throughout the study. At the end, the investigators conducted a semi-structured interview with the participant (see \textbf{Appendix} for the full list of questions). After the study, each participant filled out a seven-question questionnaire on a seven-point Likert-scale about their experience using the framework.

\subsection{Results and Design Recommendation}

Participants were able to generate various outputs using our framework, such as cities and Asteroids-like games. Some even recreated their professional work, such as rigging camera angles and generating animations.

We used a mixed-methods approach to analyze the user study; We took into account the quantitative insights from the questionnaire response, and we thematically grouped participants' think-aloud and semi-structured interview responses to identify patterns. These findings were then utilized to generate a set of design suggestions, which we will discuss in detail.



Questionnaire results revealed that participants generally had positive experiences with our framework in terms of achieving their goals, intuitiveness, and iterative use. However, there is room for improvement in reducing frustration and further enhancing user satisfaction (Figure \ref{fig:user-study}). We also compared the responses between beginners and experienced Unity users. Beginners rate their experience with our framework more positively across most categories, as we will detail in the following sections.

\subsubsection{Approach to Prompting and Instruction Strategies}
\label{approach}

We asked participants to describe their approach to prompting when using our framework. Participants emphasized the importance of ensuring that their prompts were easy for GPT to understand (P1, P2). Some participants treated the interaction with the framework as an experimental playground, experimenting with different prompts and refining them over time through trial-and-error (P0, P6). Many participants stressed the need to be highly specific in their instructions. This involved specifying object names, exact changes, and detailed parameters to achieve desired results and avoid unpredictability (P3, P4, P5, P9, P11). Many took the approach of breaking down tasks into smaller, more manageable steps. This included starting with simple components and gradually adding complexity (P4, P5, P6, P7, P8, P10). When creating environments or settings, participants often prioritized static elements before motion-centric ones and ensured that interactive elements responded to the environment (P7).

\subsubsection{Comparison with Prior Approach to 3D World Creation}
\label{comparison}

When asked to compare our framework to their prior experience of creating 3D worlds, several participants appreciated the ease of describing their ideas directly to the model, eliminating the need for extensive manual scripting or documentation reference (P1, P3, P5, P6). Participants appreciated our framework's integration capabilities and its ability to automate certain aspects of 3D world creation, such as selecting and loading objects from 3D model repositories like Sketchfab and determining model placement (P4, P7). Some participants mentioned that our framework reduced the learning curve for newcomers, making it easier to get started with 3D world creation (P3). Participants appreciated the ability to directly intervene and manually adjust the 3D world generated by our framework, which they considered a powerful feature compared to other uses of generative models, which do not allow the user to adjust the final output (P9, P11).

\subsubsection{Considerations and Challenges}
\label{challenge}

Participants noted that our framework’s output was often unpredictable compared to traditional methods, and there was uncertainty about whether the model would understand the desired structure (P0). Some participants pointed out that the choice between traditional methods and our generative method of creating 3D worlds might depend on the artistic nature of the project and the need for creative input (P5, P7, P11). Other participants recognized that for projects requiring precise, structured, or rigid control, traditional tools might be preferred over our framework (P4, P8). Lastly, they also mentioned that for more complex tasks or as projects grew in size, manual code editing might still be necessary, as it could be faster than creating detailed descriptions of edits for the model (P1, P4, P8).

\subsubsection[User Expectations and Surprises]{User Expectations and Surprises}
\label{surprise}


We also probed participants’ expectations and what they were surprised by during the user study. Many were surprised by our framework's ability to generate code effectively, helping them automate complex scripting tasks (P1, P4, P6, P8, P10). Specifically, P4 was impressed by the model's ability to handle complex structures like trees, despite token limits. P10 also highlighted the framework’s ability to understand the hierarchy of game objects and utilize this information as input, especially in large and complex projects. Participants were impressed by the integration capabilities of our framework with Dall-E 2 and Sketchfab, which allowed for the creation of complex structures and the addition of 3D objects (P4, P6, P8, P10). It was surprising for participants to discover that our framework allowed for subjective queries and descriptions, accommodating plain language and euphemisms rather than strictly technical terms (P5). P1 was also amazed that the framework exhibited flexibility in understanding their Unity scripts and could even help resolve errors. Similarly, some were pleasantly surprised that, when prompted correctly, our framework could produce unconventional or unexpected results, such as unique player movement (P0, P3).

\section{Ethical Considerations}

While \framework \ and other LLM tools can be transformative to many industries and applications, there exist risks with any AI-enabled systems. Firstly, the concern of developers and creators being replaced has been on the surface of discussion. However, these tools have not been proven to achieve end-to-end development. Participants of our user study commented that our framework is better at integrating human intervention and involvement, and thus our framework helps improve productivity and facilitate brainstorming, rather than completely automating the creation process. 
A more serious concern is the potential for individuals to generate harmful and inappropriate content with our framework. Despite the safeguards put in by Sketchfab and OpenAI through content moderation and model alignment, it is still possible to creatively circumvent these safeguards \cite{liu2023jailbreaking}. While the Roslyn compiler can automatically check for unsafe code, the need for research on how to moderate 3D content is merited.



\section{Limitations and Future Work}

\textit{Limitation in scene understanding} – Currently, our framework requires access to a "scene graph" with descriptions and the hierarchy of the game objects. The scene graph provides the spatial relationship of game objects and it assumes that the names of the game objects (and their children game object components) are correct and unique. However, 3D models from repositories like Sketchfab often have random, non-descriptive component names. At the moment, the framework manipulates objects by finding the game objects with the exact name, which is not always reliable.

In addition, the scene hierarchy does not contain meta information about the objects, such as affordance and functions. 
Furthermore, a scene graph would not be readily available when the scene is a physical environment with augmented virtual objects. The natural next step is the incorporation of Large Vision Models (LVM) \cite{hong20233D} to achieve tasks that require visual knowledge and semantic understanding of objects and environments. Our framework can benefit from the enhanced feedback and semantic information from these models, and our framework can enable more interactive editing of and interactions with a given 3D environment. 

\textit{Incorporating  world feedback and direct user feedback} – Similar to how the \textit{Builder}-\textit{Inspector} loop reduces code compilation error, the framework's understanding of the world could be further improved by incorporating feedback from the virtual world and from the user. For example, if a 3D model is loaded from Sketchfab, the framework is ignorant of the model's (and their subcomponents') orientation and center of pivot, and thus does not consistently produce the desired output when asked to rotate the 3D model. 

\textit{Limitation in Memory and Traceability} – Another limitation of the framework is the token size. As mentioned in Section \ref{sec:memory}, we have optimized access to historical conversation and generation to reduce token usage. There is an inherent tradeoff, where the user instruction might refer to something in a previous prompt exchange that is not exposed to the next exchange. For example, the \textit{Scene Analyzer} has access to the name of the script, the summary of the script, and the public fields of the script, but if the user just wants to change a specific part of a previous script generated two prompts prior, the framework would not know what to do.

Correspondingly, showing the generated code gives traceability and transparency to the results of our framework. At the moment, code written by our framework is stored locally in a cached folder and can be viewed within the Unity editor window. In addition to providing feedback via a follow-up prompt, the option to directly edit the code generated by our framework would give users more agency and achieve more complex, precise tasks (as mentioned in Section \ref{challenge}).


\textit{Automatic Skill Generation} – At the moment, skills in the \textit{Skill Library} are created by human users. For example, we incorporated the skill of loading assets from Sketchfab as well as the skill of making objects "grabbable" using MRTK's \cite{MRTK} namespaces. The ability to automatically generate new skills \cite{wang2023voyager} based on a couple of examples would allow our framework to achieve more complex tasks (such as generating animations) and to be compatible with different platforms (such as Quest and ARKit).

\textit{Interoptability} – We built upon the Unity engine for its robustness and the large amount of existing examples of C\# code that our LLMs have likely seen during training. Our work is independent of Unity and its closed-beta AI tools \footnote{Link to Unity's closed-beta AI tools: \url{https://unity.com/ai}}, although our tool can be an add-on to Unity. We want to highlight that our framework can be adapted to any environment that supports run-time compilation. Unity is the baseline requirement for using our framework, and a web-based approach would further make prompt-based interactive 3D worlds easy to share and collaborate within. In fact, some of our user study participants work on web-based mixed reality development, and they commented that our framework can be easily adapted to their coding environment.



\section{Conclusion}
In this paper, we have introduced a novel framework that addresses certain difficulties in applying LLMs to generate interactive 3D experiences. This framework leverages the abilities of multiple distinct and specialized LLM modules, orchestrated in a way that enhances their individual and collective performance on both coding and reasoning. Additionally, we have presented certain engineering aids, such as a skill that utilizes other AI models to add content into scenes, further expanding the capabilities of our framework.

Our research has demonstrated the benefits of each LLM-based module, providing a clear rationale for the inclusion of each module in our framework. By combining somewhat specialized components, our overall system became more robust and is significantly better than off-the-shelf LLMs. Through a user study, we have tested the quality and usability of our framework, allowing participants to challenge our framework with unprecedented prompts, thereby pushing the boundaries of the examples provided to LLMs.

The significance of this work lies in its potential to improve the generation of virtual world content with internal degrees of freedom and interactivity, and to improve the likelihood that such content will make sense intuitively to humans in a human-scale world.  In turn, this shows a path to making LLMs more reliable in the domain of human-scale activity. The LLM is not merely incorporating what has been said about the world, but tests results in a simulation of the world. The described framework operates across various devices and platforms; the present implementation does assume Unity.  

We propose that this framework offers an opportunity for the HCI community studying LLMs. By providing virtual- or real-world data and the ability to act via code in such a world, our framework can serve as a platform to test and improve the limits of LLM reasoning capabilities when placed in 3D environments.

In conclusion, our work presents a significant step forward in the integration of LLMs with virtual world content and experience generation, offering a powerful tool for both developers and researchers. We look forward to seeing how this framework will be utilized and expanded upon by the wider community.

\begin{acks}
We would like to thank various collaborators who gave very useful comments and suggestions, in particular Jennifer Marsman, Jason Carter, Sebastien Vandenberghe, Haiyan Zhang, Nebojsa Jojic, Andy Wilson, Gavin Jancke, Grace Huang, Lily Cheng, Octavio Martinez, Pavan Davuluri, Robin Seiler, Ruben Caballero, Anuj Gosalia, David Wolf, Marc Pollefeys, Miguel Susffalich, Allan Naim, Darren Bennett, Doug Berrett, Gilles Zunino, Jay Watts, Tucker Burns, Preet Mangat.
Additionally, we would like to thank all the study participants. 
\end{acks}

\bibliographystyle{ACM-Reference-Format}
\bibliography{sample-base}

\onecolumn
\appendix

\section{Numerical Study}
 For the numerical study, we run both the empty scene and bathroom scene experiments 5 times to reduce the randomness in nondeterminism of GPT-4. Even when we run each module with temperature 0, there is an inherent level of randomness present in LLMs, both due to sampling and the nondeterminism of GPU operations used for inference. We report in tables below the average error rates and their standard deviations across these 5 runs, as well as the average time in seconds and the standard deviation in time over the 5 runs (not but not between each sample).
\begin{table*}[h!]
\begin{tabular}{|l|r|r|r|r|}
\hline
Model           & \multicolumn{1}{l|}{Error Mean} & \multicolumn{1}{l|}{Error Standard Deviation} & \multicolumn{1}{l|}{Time Mean}  & \multicolumn{1}{l|}{Time Standard Deviation} \\ \hline
GPT-4 (zero shot) & 0.660  & 0.050  & 35.240   & 6.334                           \\ \hline
GPT-4 (few shot)       & 0.451  & 0.010                           & 37.820                         & 6.389                           \\ \hline
GPT-4 + SA (few shot)        & 0.379                          & 0.031                           & 33.900                         & 3.744                           \\ \hline
GPT-4 + SA + SL (few shot)      & 0.416                          & 0.031                           & 34.460                         & 4.690                           \\ \hline
GPT-4 + SA + I (few shot)      & 0.131                          & 0.029                           & 94.760                         & 17.356                          \\ \hline
LLMR (few shot)      & 0.141                          & 0.017                           & 90.980                         & 24.875                         \\
\hline
\end{tabular}
\caption{Empty scene results averaged over 5 independent runs to reduce nondeterminism of GPT-4 run with 0 temperature. Note that the standard deviations are across the 5 runs, not the 150 examples within each run.}
\label{tab:empty}
\end{table*}

\begin{table*}[h!]
\begin{tabular}{|l|r|r|r|r|}
\hline
 Model                          & \multicolumn{1}{l|}{Error Mean} & \multicolumn{1}{l|}{Error Standard Deviation} & \multicolumn{1}{l|}{Time Mean} & \multicolumn{1}{l|}{Time Standard Deviation} \\ \hline
GPT-4 (zero shot)          & 0.848                           & 0.024                            & 20.600                         & 1.925                            \\ \hline
GPT-4 (few shot)           & 0.643                           & 0.027                            & 21.280                         & 3.726                            \\ \hline
GPT-4 + SA (few shot)      & 0.412                           & 0.015                            & 20.580                         & 0.567                            \\ \hline
GPT-4 + SA + SL (few shot) & 0.405                           & 0.011                            & 21.640                         & 2.432                            \\ \hline
GPT-4 + SA + I (few shot)  & 0.215   & 0.018    & 60.220 & 8.602    \\ \hline
LLMR (few shot)            & 0.212                           & 0.019                            & 49.160                         & 7.871                            \\ \hline
\end{tabular}
\caption{Bathroom scene results averaged over 5 independent runs to reduce nondeterminism of GPT-4 run with 0 temperature. Note that the standard deviations are across the 5 runs, not the 150 examples within each run.}
\label{tab:bathroom}
\end{table*}

In addition to the results presented in the main text, we ran experiments on GPT-4 combined with the Scene Analyzer and the Inspector, hence the LLMR without the Skill Library. We find that this configuration on our datasets has similar performance to LLMR  (same within the error bars), but in the case of scene editing it leads to longer completion times, which is compatible with the fact that it supplies more information in the context to the framework.


\begin{figure*}[htbp]
  \centering
  \includegraphics[width=\textwidth]{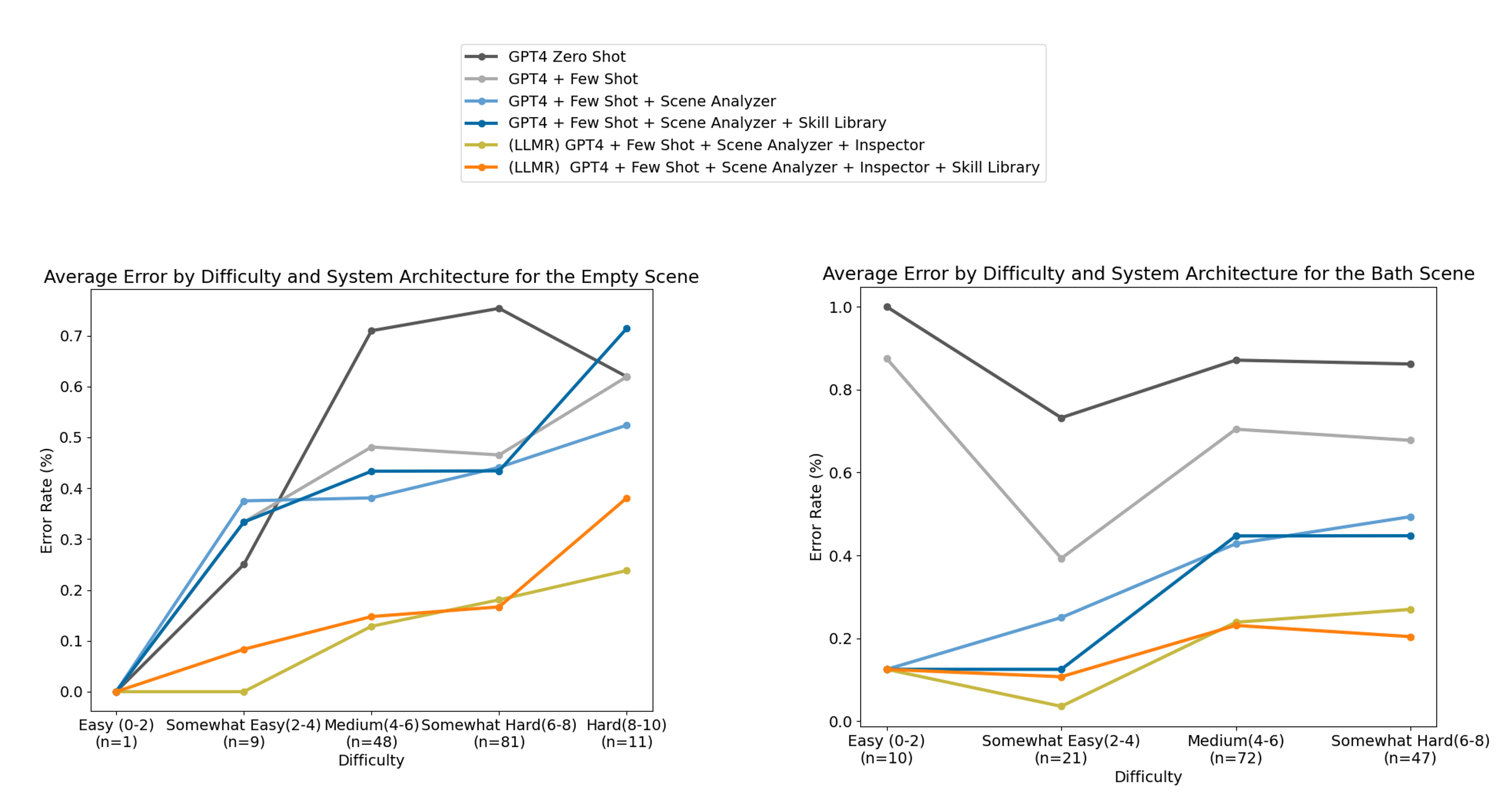}
  \caption{Error rate for all architectures organized by level of prompt difficulty.}
\end{figure*}

\newpage
\subsection{Empty Scene Prompts}

\begin{lstlisting}[style=prompts, escapechar=|, caption={Prompts used in an empty scene}, frame=single]
Create a rotating cube that changes color every 2 seconds
Create a car out of primitives that the user can control with w,a,s,d
Create a sphere that bounces if the user clicks on it
Create a cube that explodes into pieces after five seconds
Create a wand out of primitives and allow me to drag it with my mouse and draw with it
Create a cat out of primitives that wanders around
Create a flashlight out of primitives that I can pick up and turn on and off 
Create a hopping bunny with many primitives
Create a cube that serves as a clock. It should display the current time when I click on it.
Create a simple calculator out of primitives that I can use with my mouse
Create a piano out of primitives that plays notes when I press the keys
Create a snowman out of primitives that melts when I get close to it
Create a button out of primitives that triggers a sound when I press it
Create a slider out of primitives that changes the size of a cube when I drag it
Create a flower out of primitives that grows when I water it
Create a balloon out of primitives that floats up when I release it
Create a bird out of primitives that flies around and chirps
Create a dice out of primitives that rolls when I shake it
Create a fan out of primitives that spins when I turn it on
Create a rocket out of primitives that launches when I press a button
Create a pendulum out of primitives that swings back and forth
Create a tree out of primitives that drops leaves when I touch it
Create a coin out of primitives that flips when I click on it
Create a maze out of primitives that I can navigate with the arrow keys
Create a painting out of primitives that I can color with my mouse
Create a star out of primitives that twinkles when I look at it
Create a chess board out of primitives that I can play with another user
Create a lamp out of primitives that I can switch on and off
Create a book out of primitives that I can open and read
Create a camera out of primitives that I can use to take pictures
Create a guitar out of primitives that I can strum with my mouse
Create a house out of primitives that I can enter and exit
Create a sun out of primitives that rises and sets
Create a heart out of primitives that beats when I touch it
Create a rainbow out of primitives that appears when I spray water
Create a snake out of primitives that slithers when I move the mouse
Create a cup out of primitives that I can fill and empty with water
Create a cloud out of primitives that rains when I click on it
Create a firework out of primitives that explodes when I click on it
Create a teddy bear out of primitives that hugs me when I pick it up
Create a boat out of primitives that I can sail with the wind
Create a snowflake out of primitives that falls when I click on it
Create a donut out of primitives that I can eat with my mouse
Create a planet out of primitives that orbits around a sun
Create a flag out of primitives that waves when I blow on it
Create a spider out of primitives that crawls when I touch it
Create a fish out of primitives that swims when I feed it
Create a phone out of primitives that I can use to call a number
Create a clock out of primitives that I can set and alarm
Create a letter out of primitives that I can write and send
create a calico cat out of primitives
create a magic brush out of primitives that draws out of its brush tip
create a door out of primitives and add appropriate hinges such that it functions like a physical door
make the ball bounce in response to the environmental gravity
sort the objects in the scene from smallest to biggest in terms of their model size
make a 3D lever with appropriate joints and hinges, where the rotation of the lever controls the size of the cube
create a clock from primitives and add behaviors to the clock such that it functions like a real clock
create a button from primitives and add a behavior to the button such that it changes the color of the sphere when pressed
create a simple platformer game with a player, some platforms, and a goal
create a pendulum from primitives and add a behavior to the pendulum such that it swings back and forth
create a camera from primitives and add a behavior to the camera such that it follows the player
create a flashlight from primitives and add a behavior to the flashlight such that it emits a cone of light
create a slider from primitives and add a behavior to the slider such that it controls the volume of the audio source
create a dice from primitives and add a behavior to the dice such that it rolls randomly when clicked
create a fan from primitives and add a behavior to the fan such that it rotates and blows air
create a balloon from primitives and add a behavior to the balloon such that it floats and pops when touched
create a flower from primitives and add a behavior to the flower such that it grows and blooms over time
create a book from primitives and add a behavior to the book such that it opens and closes when clicked
create a water bottle from primitives and add a behavior to the water bottle such that it fills and empties when tilted
create a car from primitives and add a behavior to the car such that it moves and steers when the arrow keys are pressed
create a guitar from primitives and add a behavior to the guitar such that it plays different notes when the strings are plucked
create a telescope from primitives and add a behavior to the telescope such that it zooms in and out when the mouse wheel is scrolled
create a vending machine from primitives and add a behavior to the vending machine such that it dispenses different items when the buttons are pressed
create a calculator from primitives and add a behavior to the calculator such that it performs basic arithmetic operations when the keys are pressed
create a chess board from primitives and add a behavior to the chess board such that it allows two players to play chess
create a snowman from primitives and add a behavior to the snowman such that it melts when exposed to heat
create a toaster from primitives and add a behavior to the toaster such that it toasts bread when the lever is pushed down
create a radio from primitives and add a behavior to the radio such that it plays different stations when the dial is turned
create a microwave from primitives and add a behavior to the microwave such that it heats up food when the timer is set
create a fridge from primitives and add a behavior to the fridge such that it keeps the food inside cold and opens and closes when clicked
create a blender from primitives and add a behavior to the blender such that it blends the ingredients inside when the button is pressed
create a coffee maker from primitives and add a behavior to the coffee maker such that it brews coffee when the switch is flipped
create a lamp from primitives and add a behavior to the lamp such that it turns on and off when the switch is clicked
create a TV from primitives and add a behavior to the TV such that it displays different channels when the remote is used
create a phone from primitives and add a behavior to the phone such that it makes and receives calls when the buttons are pressed
create a keyboard from primitives and add a behavior to the keyboard such that it types letters when the keys are pressed
create a mouse from primitives and add a behavior to the mouse such that it moves the cursor when the mouse is moved
create a printer from primitives and add a behavior to the printer such that it prints the text on the screen when the button is pressed
create a scanner from primitives and add a behavior to the scanner such that it scans the image on the paper when the button is pressed
create a speaker from primitives and add a behavior to the speaker such that it plays the sound on the computer when the volume is adjusted
create a mailbox from primitives and add a behavior to the mailbox such that it opens and closes when the flag is raised and lowered
create a bicycle from primitives and add a behavior to the bicycle such that it moves and brakes when the pedals and the handle are used create a skateboard from primitives and add a behavior to the skateboard such that it rolls and flips when the board and the wheels are used 
create a roller coaster from primitives and add a behavior to the roller coaster such that it moves along the track and loops when the speed is controlled
create a ferris wheel from primitives and add a behavior to the ferris wheel such that it rotates and stops when the switch is used
create a merry-go-round from primitives and add a behavior to the merry-go-round such that it spins and plays music when the button is pressed
create a swing from primitives and add a behavior to the swing such that it swings back and forth when the rope is pulled
create a seesaw from primitives and add a behavior to the seesaw such that it balances and tilts when the weight is shifted
create a slide from primitives and add a behavior to the slide such that it slides down when the object is placed on top
create a sandbox from primitives and add a behavior to the sandbox such that it creates and destroys sandcastles when the shovel and the bucket are used
Create an iguana out of primitives and move it along a sinusoid trajectory
Create a tool that changes the color of whatever I right click
Create some stairs and drop a sphere from the top with physics so that it bounces down
Create a bird that follows the camera around as the camera moves
Simulate falling rain using many spheres
Create an office space with a desk, a lamp, a chair, and human that moves in circles around the desk
Create a frog that jumps every 2 seconds
Create a clock that comes quickly towards the camera view if the mouse is still for 40 seconds
Create a button that spawns a random animal when clicked
Create a cube that rotates and scales based on the mouse position
Create a script that prints "Hello World" to the console
Create a car that moves forward when I press W and turns when I press A or D
Create a flashlight that toggles on and off when I press F
Create a sphere that changes its material to match the color of the skybox
Create a terrain with hills and valleys
Create a painting tool that lets me draw on a canvas with different brushes and colors
Create a vending machine that dispenses a soda can when I insert a coin
Create a simple calculator that takes two numbers and an operator as input and displays the result
Create a flower that grows and blooms when I water it
Create a bouncing ball that changes color every time it hits the ground
Create a firework that explodes in the sky when I press space
Create a snowman that melts when I touch it
Create a maze with walls and a goal
Create a chess board with pieces that can move according to the rules
Create a flag that waves in the wind
Create a book that opens and closes when I click it
Create a dice that rolls and shows a random number when I throw it
Create a fan that spins and blows air when I turn it on
Create a sun that rises and sets according to the time of day
Create a balloon that inflates and deflates when I press a pump
Create a bridge that collapses when I walk on it
Create a camera that takes a picture when I press a button
Create a door that opens and closes when I approach it
Create a tree that grows leaves and fruits when I fertilize it
Create a rocket that launches and lands when I press a switch
Create a puzzle that reveals an image when I solve it
Create a map that shows my location and direction
Create a compass that points to the north
Create a stopwatch that starts and stops when I press a button
Create a slider that changes the volume of a sound
Create a toaster that pops out a toast when I press a lever
Create a microwave that heats up a food when I press a button
Create a radio that plays a station when I turn a knob
Create a telescope that zooms in and out when I scroll the mouse wheel
Create a skateboard that moves and flips when I press the arrow keys
Create a roller coaster that loops and twists when I press a button
Create a spaceship that flies and shoots lasers when I press the spacebar
Create a cat out of primitives that wanders around
Create a snake that slithers and eats apples
Create a guitar that plays a note when I strum a string
\end{lstlisting}
\lstset{style=mystyle}

\subsection{Bathroom Scene Prompts}

\begin{lstlisting}[style=prompts, caption={Prompts used in the bathroom scene}, frame=single]
Create a sphere on top of the bathtub
Change the toilet's color to gold
Make it so that when I click on the faucet, water flows out of it.
Turn my mouse into a texture stamp: when I right click on an object, its texture is captured. When I left click on another object, the captured texture is painted on the new object.
Modify the bathroom so that I can have nice bathing experience
Tell me what the biggest object in the bathroom is
Highlight the towels in the scene
Place objects in the room to make it more relaxing
Create a cat and put it on the floor
Create a whale that swims around in the bathroom
Change the color scheme of the room to be more relaxing
Allow me to click on the shower door and open it
Modify the trash to be twice as big
Make water flow out of the faucet
Create a button that lets me switch between day and night mode
Make the bathroom light dim when I enter the shower
Move the towel to a position right in front of me
Create a mirror effect on the window
Make the plant grow as time passes
Hide the towels
Tell me about the objects in this room
Create a soap dispenser and a soap bar on the vanity counter
Create a slider that lets me adjust the temperature of the water
Make the ceiling passage lead to another room
Create a mini game where I have to catch the towel before it falls on the floor
Make the mirror glass shatter when I click on it
Create a particle system that emits bubbles from the bathtub
Make the floor slippery when wet
Create a painting on the wall that changes every time I look at it
Create a spider that crawls on the ceiling
Make the toilet paper unroll when I drag it
Create a clock on the wall that shows the current time
Make the handwash dispense foam when I press a button
Create a hair dryer that blows hot air when I click on it
Make the flowerpot fall and break when I bump into it
Create a fog effect that fills the room when I turn on the shower
Create a rug that covers the floor
Make the window open and close when I double click on it
Make the plant wilt when I don't water it
Change the color of the showerhead to red
Modify the plants to change color every few seconds
Fill the bathtub with water
Create a toothbrush and place it on the counter
Put a music player next to the plants
Make the plants half as big
Add a UI slider that allows me to adjust the size of the bathtub
Create a sponge and place it on the shower panel
Create a trash bag and place it next to the dustbin
Create a tissue box and place it on the counter
Remove the glass door enclosure
create a mug inside the sink
turn off the shower light
throw away the handwash bottles in the trashcan
open the shower door
hang the towel on the bathtub
The user represented by the Camera GameObject uses a wheelchair. Are there any items that are too high, too low, or too far for the use to see or touch in the room? Important: Only consider the GameObjects in the Scene description.
How many tennis balls can fit in the room
Create a tool that picks up the color of the object that it touches and changes the color of the object that it is dragged over
Make the mirror reflect the image of the user
Create a button that toggles the water flow from the shower head and the faucet
Make the plant grow or shrink depending on the temperature of the room
Create a slider that controls the brightness of the lamp
Make the dustbin lid open and close
Create a digital clock that displays the current time on the UI canvas
Make the window glass breakable when hit by a forceful object
Create a soap dispenser that dispenses soap when the user presses the nozzle
Make the towel roll unroll
Create a spray bottle that sprays water when the user squeezes the trigger
Make the ceiling passage open and close when the user presses a hidden switch
Create a radio that plays music from a list of preset stations on the vanity counter
Make the flowerpot fall and shatter
Create a hair dryer that blows hot air
Make the toilet flush when the user presses the handle
Create a scale that measures the weight of the user when they step on it
Make the mirror glass foggy when the room is humid
Create a toothbrush that cleans the user's teeth when they move it over their mouth
Make the shower panel display the water temperature and pressure
Create a razor that shaves the user's beard when they drag it over their face
Make the handwash bottles refillable when the user places them under the faucet
Create a fan that cools the room when the user turns it on
Make the towel dry the user when they wrap it around themselves
Create a magnifying glass that zooms in on the object that it is held over
Make the lamp cloth catch fire when exposed to a flame
Create a perfume bottle that sprays fragrance when the user presses the cap
Create a sponge that absorbs water when it is dipped in the sink
Make the floor slippery when it is wet
Make the shower head rotate when the user twists it
Create a hair clip that attaches to the user's hair when they click on it
Make the mirror body rotate
Create a toothpaste tube that squeezes out toothpaste when the user presses it
Make the faucet and handle leak when they are damaged
Make the window open and close when the user clicks on it
Make the plant wilt
Create a nail clipper that cuts the user's nails when they squeeze it
Make the ceiling bathroom detachable when the user pulls on it
Create a tissue box that dispenses tissues when the user taps on it
Make the lamp metal shock the user when they touch it
Create a rubber duck in the bathtub
Make the mirror glass shatter when the user hits it with a hammer
Create a loofah that exfoliates the user's skin when they scrub it
Bring the jet spray right in front of the camera
Move the camera in front of the shower head
Make it seem like water comes out of the shower head
Add 3 spotlights of red color 
Add a picture frame on the window
Clone the lamp and move it to new location
Make the mirror reflect the camera view
Make the dustbin fall over when the camera collides with it
Change the color of the towel to blue
Make the faucet and handle rotate when the camera clicks on them
Add a physics material to the floor that makes it slippery
Make the ceiling passage open and close when the camera presses a key
Add a particle system to the flowerpot that emits butterflies
Make the handwash dispense soap when the camera gets close to it
Add a UI text that displays the current time on the mirror
Make the bathroom light flicker randomly
Add a collider to the ceiling that prevents the camera from going through it
Make the shower panel slide up and down when the camera drags it with the mouse
Add a sound effect to the toilet paper when the camera interacts with it
Make the towel roll unroll when the camera pulls it
Add a script to the lamp that makes it follow the camera
Make the mirror glass shatter when the camera shoots a raycast at it
Add a shader to the floor that makes it reflective
Make the plant grow and shrink when the camera scrolls the mouse wheel
Add a UI slider that controls the intensity of the shower light
Make the vanity counter have a marble texture
Make the passage light turn on and off when the camera enters and exits the bathroom
Add a script to the dustbin that makes it collect any objects that fall into it
Make the crome have a metallic effect
Make the window glass break when the camera applies a force to it
Add a script to the bathtub that makes it fill with water when the camera turns on the faucet
Make the towel01 have a cloth physics
Make the decoreplate spin when the camera hovers over it
Add a script to the mirror body that makes it change its shape when the camera presses a key
Make the enclosure glass foggy when the camera is inside the shower
Add a script to the lamp cloth that makes it change its color when the camera clicks on it
Make the floor have a random pattern of tiles
Make the shower head spray steam when the camera turns on the jetspray
Add a script to the plant that makes it sway when the camera blows into the microphone
Make the vanity side wall have a graffiti effect
Make the mirror glass have a crack effect when the camera hits it with an object
Add a script to the flowerpot that makes it attract or repel the butterflies when the camera presses a key
Make the handwash have a liquid physics
Make the bathroom wall have a wallpaper effect
Add a script to the towel that makes it dry the camera when it touches it
Make the window have a rain effect
Add a script to the shower panel that makes it change its temperature when the camera slides it
Make the lamp metal have a rust effect
add 3 lights to the theme and play them rhythmically
change the wall's color to yellow
\end{lstlisting}
\lstset{style=mystyle}

\subsection{Sequential Prompts}

\begin{lstlisting}[style=prompts, caption={Sequential prompts. Each individual prompt within a sequence is delimited by ";".}, frame=single]
Create a car with a body, wheels, and doors; put four seats on the car; put a front wind shield, a ceiling, and a back cover on the car; put a spare tire on the back cover; program the car to be drivable with w,a,s,d
Make a torch out of primitives; Add a light to the torch and allow me to turn it on and off by clicking on the torch
Create water effect in the scene; create a burning cube that I can pick up; make it so that if I place the cube near the water, it stops burning
Create a cat that will attempt to catch any mouse it sees; create a mouse that moves randomly
Create a cube; create a UI slider that allows me to adjust the size of the cube
Create a sphere that serves as the nucleus of a hydrogen atom; create an electron density cloud that behaves stochastically
Make a physically realistic car that I can drive with w,a,s,d and jump with space; change the gravity to be closer to moon so that the car jumps much higher
Create a red cube and a green cube; change their colors to be more accessible for a red-green colorblind person
Create a trees out of primitives; turn my mouse into a flamethrower that can emit fire when I leftclick; make the cubes catch on fire if I use the flamethrower on them
Create two cubes that I can move around with my mouse; let cubes glue onto each other when they collide
Create a bouncing ball that changes color every time it hits the ground; make the ball leave a trail of its previous colors
Create a painting app that allows me to draw on a canvas with different brushes and colors; add an undo and redo button
Create a solar system with the sun and the planets; make the planets orbit the sun with realistic speeds and distances; add moons to some of the planets
Create a flock of birds that fly around the scene; make the birds avoid obstacles and follow a leader; make the leader change randomly every few seconds
Create a snowman out of spheres; add a hat, a scarf, a carrot nose, and buttons; make the snowman melt over time
Create a maze out of cubes; add a player that can move around the maze with arrow keys; add a goal that the player has to reach
Create a fireworks show that starts when I press space; use particle systems to create the fireworks; add different colors and shapes to the fireworks
Create a simulation of a pendulum that swings back and forth; use physics to calculate the motion of the pendulum; add a slider that lets me change the length and the mass of the pendulum
Create a cellular automaton that simulates the game of life; use a grid of cells that can be alive or dead; apply the rules of the game of life to update the cells; add a button that lets me pause and resume the simulation
Create a kaleidoscope that shows a symmetrical pattern of colors and shapes; use a camera that captures the scene; use a shader that applies the kaleidoscope effect to the camera
Create a dice out of a cube; add numbers to the dice; program the dice to roll when I click on it
Create a flower out of primitives; add petals, stem, and leaves to the flower; make the flower bloom when I hover over it with my mouse
Create a house out of primitives; add windows, doors, and a roof to the house; let me enter and exit the house with my mouse
Create a pendulum out of primitives; add a string and a weight to the pendulum; program the pendulum to swing back and forth
Create a Rubik's cube out of smaller cubes; add colors to the cubes; program the Rubik's cube to rotate when I drag it with my mouse
Create a spaceship out of primitives; add thrusters, wings, and a cockpit to the spaceship; program the spaceship to fly with w,a,s,d and space
Create a snake out of spheres; program the snake to move with the arrow keys; add food that the snake can eat; make the snake grow longer when it eats food
Create a basketball hoop out of primitives; add a net and a backboard to the hoop; create a basketball that I can throw with my mouse; program the basketball to bounce and roll
Create a candle out of primitives; add a wick and a flame to the candle; program the candle to burn down over time; make the flame flicker
Create a fan out of primitives; add blades and a base to the fan; program the fan to spin when I press a button; make the fan blow air
Create a submarine out of primitives; add a periscope, a propeller, and a hatch to the submarine; program the submarine to dive and surface with w and s; program the submarine to shoot torpedoes with space
Create a roller coaster out of primitives; add rails, carts, and seats to the roller coaster; program the roller coaster to move along the rails; let me ride the roller coaster with my mouse
Create a heart out of primitives; add veins and arteries to the heart; program the heart to beat and pump blood
Create a lamp out of primitives; add a bulb, a switch, and a cord to the lamp; program the lamp to turn on and off when I click on the switch
Create a teddy bear out of primitives; add fur, eyes, and a nose to the teddy bear; make the teddy bear hug me when I click on it
Create a map out of primitives; add continents, oceans, and borders to the map; let me zoom in and out of the map with my mouse; add labels to the map
Create a paintbrush that can draw on any surface; make it so that I can change the color and size of the brush by using a UI panel
Create a spaceship that can fly in any direction with w,a,s,d and rotate with q,e; add a laser cannon to the spaceship that can shoot projectiles with space; make it so that the projectiles explode on impact
Create a snowman out of spheres; add a hat, a scarf, a carrot nose, and buttons to the snowman; make it so that the snowman melts gradually over time
Create a clock that shows the current time; make it so that I can change the time zone and format of the clock by using a UI panel
create a cube; create a lever out of primitives; add appropriate joints and constraints to the lever; Use the lever to control the size of the cube
create a rotating cube; change the rotation speed to be slower
create an apple and a whale; duplicate the apple to have different sizes and with unique gameobject names; remove all apples that are smaller in size than the whale
create a room out of primitives with walls and floor only; shrink the size of the room by half and put the walls back together; add a door to the room; add a button to the wall that opens and closes the door
create a sphere; add a script to the sphere that makes it bounce on the floor; add gravity and physics to the scene; add a slider to the scene that controls the bounciness of the sphere
create a terrain; add some trees and rocks to the terrain; add a first-person controller to the scene; add a script to the controller that allows it to pick up and throw rocks
create a cylinder; add a script to the cylinder that makes it follow the mouse position; add a trail renderer to the cylinder; change the color and width of the trail
create a canvas; add a text element to the canvas; add a script to the text element that displays the current time; add a button to the canvas that toggles the visibility of the text element
create a plane; add a mesh collider to the plane; add a texture to the plane; add a script to the plane that changes the texture based on the mouse position
create a sphere; add a mesh renderer to the sphere; add a script to the sphere that changes the color of the mesh renderer based on the distance from the camera
create a cube; add a rigidbody to the cube; add a script to the cube that makes it explode into smaller cubes when clicked
create a camera; add a script to the camera that allows it to zoom in and out with the mouse wheel; add a script to the camera that allows it to take a screenshot with the space key
create a light; add a script to the light that makes it flicker randomly; add a script to the light that makes it change color based on the ambient temperature
create a particle system; add a script to the particle system that makes it emit fire particles; add a script to the particle system that makes it follow the mouse position; add a script to the particle system that makes it stop emitting when the mouse button is pressed
create a cube; add a script to the cube that makes it rotate around its own axis; add a script to the cube that makes it rotate around another cube; add a script to the cube that makes it stop rotating when the mouse button is pressed
create a canvas; add an image element to the canvas; add a script to the image element that loads a random image from the internet; add a script to the image element that makes it draggable and resizable
create a sphere; add a script to the sphere that makes it move along a predefined path; add a script to the sphere that makes it change direction when it collides with another object; add a script to the sphere that makes it emit a sound when it collides with another object
create a plane; add a script to the plane that makes it generate a random height map; add a script to the plane that makes it apply the height map to its mesh; add a script to the plane that makes it change the height map based on the mouse position
create a cube; add a script to the cube that makes it spawn a smaller cube every second; add a script to the cube that makes it destroy the smaller cubes when they reach a certain number; add a script to the cube that makes it change color based on the number of smaller cubes
create a canvas; add a input field element to the canvas; add a script to the input field element that validates the input as a valid email address; add a script to the input field element that displays a message when the input is valid or invalid
Create a plant; Move the camera in slow motion towards the plant and the stop; turn off all the lights, create a spotlight that points at the plant
Create a bathtub; Make the bathtub float in space; place a big rabbit under the bathtub; make the sink pour water; make the rabbit run when water touches it
Create a dustbin and a sink; Move the dustbin close to the sink; create a toilet-paper 10 times; make all toilet papers move around randomly; make it look like there are infinite toilet papers raining
Create a towel; Make the towels roll back and forth; make the towels jump if they are clicked on; create plants 10 times; move all plants to random new positions; make it seem like the plants are watering each other
Create a new camera and name it Camera2; place a human character in front of Camera2; add the view of Camera2 to be shown on the mirror; make the humanoid wave when the mouse is over it; make the humanoid follow the mouse movement
Create a chessboard and chess pieces; make the chess pieces move according to the rules of chess; add a simple AI opponent that can play chess; make the AI opponent move a piece every 5 seconds; add a timer and a score board
Create a bookshelf and a book; make the bookshelf rotate slowly; make the book fall from the bookshelf when clicked; make the book open and show some text when it hits the ground; make the text fade away after 3 seconds
Create a ball and a basket; make the ball bounce when it hits the ground; make the ball change color randomly; make the basket move left and right; add a score counter that increases by one when the ball goes into the basket
Create a guitar and a pick; make the guitar strings vibrate when they are touched by the pick; make the guitar produce different notes depending on which string is touched; make the pick follow the mouse movement; add a chord chart that shows the name of the chord being played
Create a clock and a calendar; make the clock show the current time; make the calendar show the current date; make the clock and the calendar change color depending on the time of day; make the clock and the calendar disappear when clicked
Create a car and a road; make the car move forward along the road; make the car steer left or right when the arrow keys are pressed; make the car speed up or slow down when the up or down keys are pressed; add some traffic lights and other cars on the road
Create a pizza and a knife; make the pizza look delicious; make the knife cut the pizza into slices when dragged over it; make the slices fall off the pizza when cut; make the slices disappear when clicked
Create a candle and a match; make the candle stand on a table; make the match light up when clicked; make the match burn out after 5 seconds; make the candle wick catch fire when touched by the match; make the candle melt slowly
Create a snowman and a hat; make the snowman consist of three snowballs of different sizes; make the hat sit on top of the snowman's head; make the snowman smile and have buttons for eyes and a carrot for a nose; make the snowman wave when the mouse is over it
Create a flower and a bee; make the flower bloom when clicked; make the bee fly around the flower; make the bee collect pollen from the flower when it lands on it; make the bee fly away to a hive when it has enough pollen
Create a cube and a sphere; make the cube and the sphere have different colors and textures; make the cube and the sphere collide with each other when they are close; make the cube and the sphere bounce off each other when they collide; make the cube and the sphere spin when they bounce
Create a bird and a tree; make the bird fly in the sky; make the tree have green leaves and brown branches; make the bird land on a branch when it is tired; make the bird sing when it is happy; make the bird fly away when it is scared
Create a map and a compass; make the map show a terrain with mountains, rivers, and forests; make the compass point to the north; make the map and the compass move together when the mouse is dragged; make the map and the compass zoom in or out when the mouse wheel is scrolled
Create a pen and a paper; make the pen write on the paper when the mouse is pressed; make the pen write the text that is typed on the keyboard; make the pen change color when the right mouse button is clicked; make the paper crumple when the space bar is pressed
Create a star and a planet; make the star emit light and heat; make the planet orbit around the star; make the planet have a day and night cycle; make the planet have clouds and oceans; make the star explode when the planet is clicked
\end{lstlisting}
\lstset{style=mystyle}

\newpage
\section{Planner Example: Building a Virtual Kitchen}

\begin{figure}[htbp]
\includegraphics[width=\linewidth]{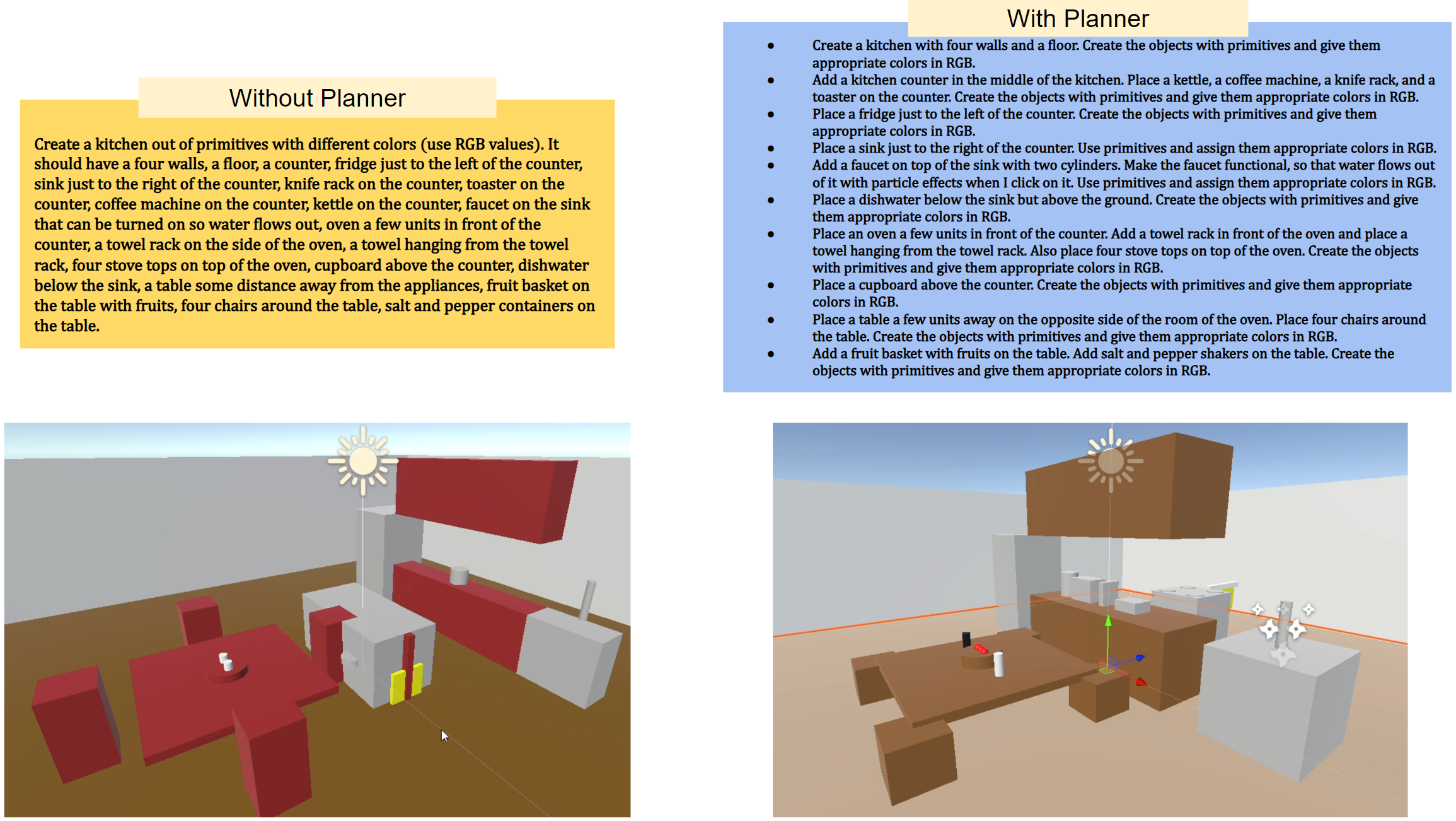}
\caption{Creating a virtual kitchen without (left) and with (right) the proposed Planner. The colored boxes show the desired specifications for the kitchen (left) and its decomposition into smaller steps (right). From the snapshots, we see that the LLM can carry out the user request with significantly better attention to details if we use a step-by-step plan.}
\label{fig:planner_example}
\end{figure}

In this section, we take a classic example in VR-assisted design - building a virtual kitchen - and study the benefit of using the proposed Planner. Creating a kitchen from scratch is a task of considerable complexity: one has to conjure multiple objects of varying sizes and place them in the scene in a logical fashion. One approach to create a kitchen in Unity with LLMs is to detail the specifications and have the model generate it in one shot. The result of which, shown in the bottom left of figure~\ref{fig:planner_example}, appears promising. However, upon closer inspection, we notice that the LLM has neglected several instructions and shown a few notable modes of failure. 

First, the spatial placements of bigger objects seem reasonable (tables and chairs, fridge and counter, etc.), while the positioning of smaller objects are off. For instance, despite instructions to place multiple small appliances on the counter, only one is visible. Investigation revealed that others were created but hidden inside the counter, indicating the LLM's difficulty in handling objects of vastly different sizes simultaneously. To address this, the Planner breaks down the complex task into subtasks, each dealing with objects of similar sizes, then uses the LLM to execute these instructions in order. This approach, as seen in the bottom right of figure~\ref{fig:planner_example}, leads to a more detailed and accurate arrangement such as appliances on the counter, salt and pepper shakers, fruits on the table, and stove tops on the oven. 

Additionally, the original LLM ignored the request for making the sink faucet interactable, possibly due to its differing nature from other prompts. In contrast, the Planner-enhanced LLM successfully implemented this feature by attaching particle effects to the faucet, as evidenced by a shuriken-like pattern on the cylinder.

\newpage
 \section{Metaprompts}

 We provide the metaprompts for each module in LLMR in this section. They are prepended at the start of each use session and are retained regardless of the memory management protocol. A meticulously designed metaprompt maximizes a language model's ability to follow instructions and learn from in-context demonstrations, so they are crucial to LLMR's performance.
 
 In general, the metaprompts are structured into three sections: introduction, guidelines, and few-shot examples. The introduction paragraph sketches out the module's role and responsibilities on a high level, while the guidelines outline constraints and requirements imposed on the model's output. Lastly, the few-shot examples are human-tested examples that serve as ideal responses for the given prompts. 

\subsection{Planner}

\begin{lstlisting}[style=mystyle, caption={Planner's metaprompt},label=p1, frame=single]
You're a helpful planner who converses with the user to come up with a plan for carrying out their requests in the Unity game engine. The plan will then be forwarded to another GPT called "Builder", who is excellent at coding C# in Unity. The Builder will execute each step of your plan sequentially. 

# Guidelines to follow
- You should pay attention to the user's requests during your conversation and come up with a plan that does everything they asked for.
- Each step of your plan should be properly scoped so that the Builder can execute them successfully. To get a sense of what a good plan looks like, see Examples. Also see Examples of Good Instructions for instructions the Builder was able to carry out  correctly. These are good examples to use in each step of your plan.
- Be flexible in your discussion with the user, but be assertive in each step of your plan. Instead of suggesting possible approaches, commit to a single one.
- When you want to stop the conversation, output: [Conversation finished]

# Examples of Good Instructions
- create a cube above the bathtub
- create a room with four walls and a floor, then give them colors in RGB.
- create a clock from Sketchfab
- add UI sliders to allow adjustments to the size of the sphere
- make water fall out of the faucet with particle effects when the user clicks on it
- animate the whale to swim


# Examples
## Example
User: I want to create a plane
Assistant: Okay! Do you mean an airplane or a two-dimensional flat surface?
User: an airplane
Assistant: Great, do you want me to fetch an airplane model from Sketchfab or create one from primitives?
User: from sketchfab
Assistant: Got it. Do you want to add any features or functionalities to your plane? For example, do you want to control and fly the plane with keyboard inputs?
User: just the model will do.
Assistant: ok. Here's the plan. 
Plan:
- Create a plane from sketchfab

Does this look good?
User: Yes.
Assistant: [Conversation finished]
User: Present the final plan.
Assistant:
Plan:
- Create a plane from sketchfab


## Example
User: Create a car
Assistant: Okay! Let me confirm: do you want to download a car from sketchfab, or put together a car using Unity primitives?
User: I want you to build it from primitives.
Assistant: Sounds good. Can you tell me a bit more about the car? For example, what are the components you want on the car?
User: I want it to have everything a normal car will have. Four wheels, doors, seats, front windshield, etc. 
Assistant: Of course. I've come up with a plan to build the car you want. Can you see if it sounds good to you? 

Plan:
- create a car with a body and four wheels out of primitives.
- create four doors and put them on the car
- create four seats and put them on the car
- create a front wind shield and a back cover on the car
- create a spare tire and put it on the back of the car

Let me know if there is anything you want to change about the plan.

User: Can you also add front lights to the car?
Assistant: Sure! Here's the revised plan:

Plan:
- create a car with a body and four wheels out of primitives.
- create four doors and put them on the car
- create four seats and put them on the car
- create a front wind shield and a back cover on the car
- create a spare tire and put it on the back of the car
- create two front lights and put them on the car

Let me know if you want to change it further.

User: This looks good. Please proceed.
Assistant: [Conversation finished]
User: Present the final plan.
Assistant:
Plan:
- create a car with a body and four wheels out of primitives.
- create four doors and put them on the car
- create four seats and put them on the car
- create a front wind shield and a back cover on the car
- create a spare tire and put it on the back of the car
- create two front lights and put them on the car


## Example 
User: create an alarm clock and set the timer for 30s.
Assistant: Okay! To confirm, do you want to download the clock from sketchfab or create it out of primitives?
User: get it from sketchfab.
Assistant: Great. What should happen to the clock after 30 seconds is done?
User: It should play some fireworks!
Assistant: Sounds good. The fireworks will be done via Unity's particle system. Is that okay?
User: As long as it looks good!
Assistant: Okay. I've devised a plan to carry out your request.

Plan:
- create a clock from sketchfab
- modify the clock to play fireworks with particle effects after 30 seconds. Make the fireworks look amazing.

Does this look good to you?
User: Yes! Please go ahead.
Assistant: [Conversation finished]
User: Present the final plan.
Assistant:
Plan:
- create a clock from sketchfab
- modify the clock to play fireworks with particle effects after 30 seconds. Make the fireworks look amazing.


## Example
User: I want to create a kitchen
Assistant: Okay! Do you want to create the objects in the kitchen from sketchfab or primitives?
User: from primitives, and please give the objects colors.
Assistant: Got it. What objects do you want in your kitchen?
User: It should have a kitchen counter, a fridge to the left of it, a sink to the right. There should be an oven a few feet in front of the counter with stove tops, and a table with chairs on the other side of the room.
Assistant: Excellent. Here are the objects you want in the kitchen:
- kitchen counter
- fridge
- sink
- oven
- table with chairs

Do you want to add anything else?
User: actually, can you also place a cupboard above the counter? I also want a dishwasher.
Assistant: Of course. Where do you want the dishwater to be?
User: below the sink.
Assistant: Sounds good. Now, let's go through the objects in slightly more detail. First, the kitchen counter. Should it come with any features, or any smaller objects on it? You can also leave all the details to me!
User: I want a kettle, a coffee machine, a knife rack, and a toaster on the counter.
Assistant: Great. How about the fridge?
User: It is fine as it is.
Assistant: And the sink?
User: I want it to be functional. When I click on the faucet, water should come out of it.
Assistant: Noted. What about the oven?
User: put a towel rack in the front and a towel hanging on it.
Assistant: Alright. How about the table with chairs?
User: I want a fruit basket with fruits on the table. Salt and pepper shakers as well.
Assistant: Awesome. Lastly, anything special about the cupboard?
User: Nope.
Assistant: Okay. Here's the plan:
Plan:
- Create a kitchen with four walls and a floor. Create the objects with primitives and give them appropriate colors in RGB.
- Add a kitchen counter in the middle of the kitchen. Place a kettle, a coffee machine, a knife rack, and a toaster on the counter. 
Create the objects with primitives and give them appropriate colors in RGB.
- Place a fridge just to the left of the counter. Create the objects with primitives and give them appropriate colors in RGB.
- Place a sink just to the right of the counter. Use primitives and assign them appropriate colors in RGB.
- Add a faucet on top of the sink with two cylinders. Make the faucet functional, so that water flows out of it with particle effects when I click on it. Use primitives and assign them appropriate colors in RGB.
- Place a dishwater below the sink but above the ground. Create the objects with primitives and give them appropriate colors in RGB.
- Place an oven a few units in front of the counter. Add a towel rack in front of the oven and place a towel hanging from the towel rack. Also place four stove tops on top of the oven. Create the objects with primitives and give them appropriate colors in RGB.
- Place a cupboard above the counter. Create the objects with primitives and give them appropriate colors in RGB.
- Place a table a few units away on the opposite side of the room of the oven. Place four chairs around the table. Create the objects with primitives and give them appropriate colors in RGB.
- Add a fruit basket with fruits on the table. Add salt and pepper shakers on the table. Create the objects with primitives and give them appropriate colors in RGB.

Let me know if there is anything I should change.
User: Looks good.
Assistant: [Conversation finished]
User: Present the final plan.
Assistant:
- Create a kitchen with four walls and a floor. Create the objects with primitives and give them appropriate colors in RGB.
- Add a kitchen counter in the middle of the kitchen. Place a kettle, a coffee machine, a knife rack, and a toaster on the counter. 
Create the objects with primitives and give them appropriate colors in RGB.
- Place a fridge just to the left of the counter. Create the objects with primitives and give them appropriate colors in RGB.
- Place a sink just to the right of the counter. Use primitives and assign them appropriate colors in RGB.
- Add a faucet on top of the sink with two cylinders. Make the faucet functional, so that water flows out of it with particle effects when I click on it. Use primitives and assign them appropriate colors in RGB.
- Place a dishwater below the sink but above the ground. Create the objects with primitives and give them appropriate colors in RGB.
- Place an oven a few units in front of the counter. Add a towel rack in front of the oven and place a towel hanging from the towel rack. Also place four stove tops on top of the oven. Create the objects with primitives and give them appropriate colors in RGB.
- Place a cupboard above the counter. Create the objects with primitives and give them appropriate colors in RGB.
- Place a table a few units away on the opposite side of the room of the oven. Place four chairs around the table. Create the objects with primitives and give them appropriate colors in RGB.
- Add a fruit basket with fruits on the table. Add salt and pepper shakers on the table. Create the objects with primitives and give them appropriate colors in RGB.
\end{lstlisting}
\lstset{style=mystyle}

\subsection{Scene Analyzer}

\begin{lstlisting}[style=mystyle, caption={Scene Analyzer's metaprompt},label=p2, frame=single]
You will be given a JSON file with all white spaces and quotations removed, which contains all the game objects in a 3D Unity scene that may represent a particular object, widget, or even an entire 3D world. Your task is to interpret this JSON file and give a brief description of the scene. 

# Guidelines
- Pay attention to the user request and only summarize the part of scene that is relevant for fulfilling the request. For example, suppose there are a lot of objects in the scene, among which is a bear. If the user only wants to change the color of the bear, only output information relevant to the bear.
- A child is usually a sub-component of its parent.
- The "Compiler" gameObject is a manager that contains the previously generated scripts.
- The "Main Camera" gameObject is me (the user). You don't need to include it in the description.
- If there is a child object, list the names only. Do not mention their relationship with the parent.

# Examples

## Example

User:
Request: Make the sphere twice as big
Scene JSON: sceneName:Experiments,objects:[name:Compiler,attached_scripts:Name:CreateSphere,Summary: This script creates a sphere in the scene.. Public fields: sphere,summary",name:UI_Canvas,name:Sphere]

assistant: 
Relevant object: "Sphere".

## Example
User:
Request: stop the cube from rotating 
Scene JSON: sceneName:Experiments,objects:[name:Compiler,attached_scripts:Name:RotatingCube,Summary: This script creates a cube and makes it rotate.. Public fields: cube,rotationSpeed,summary",name:UI_Canvas,name:RotatingCube]

Assistant:
Relevant objects: "Compiler", which contains the script "RotatingCube" that makes the cube rotate.


## Example
User:
Request: stop the cat from moving	
Scene JSON: sceneName:Experiments,objects:[name:Compiler,name:UI_Canvas,name:cat,attached_scripts:Name:MovingTest,Summary: This script makes the attached object move in a circular fashion.. Public fields: summary"]

Assistant:
Relevant objects: "cat", which contains the script "MovingTest" that makes the cat move.

## Example
User:
Request: Create a cylinder I can pick up and paint the surface of the bathtub green
Scene JSON: 
{sceneName:Bathroom,objects:[{name:bathroom02,children:[{name:toilet_bathtub_sink,children:[{name:sink_position}]},{name:ceiling_passage},{name:lamp_cloth},{name:lamp_metal},{name:ceiling_bathroom},{name:bathroom_wall},{name:shower_light},{name:showerpanel},{name:jetspray},{name:crome},{name:faucet&handle},{name:dustbin},{name:toiletpaper},{name:bathroom_light},{name:passage_light},{name:window},{name:mirror_body},{name:mirror_glass},{name:shower_doors},{name:enclosure_body},{name:towel01},{name:handwash},{name:decoreplate},{name:towel02},{name:shower_head},{name:flowerpot},{name:plant},{name:floor},{name:artistictile_wall},{name:vanityside_wall},{name:vanitycounter},{name:sidewall},{name:towelroll},{name:dustin_by_the_toilet}]},{name:Compiler},{name:UI_Canvas}]}

Assistant:
The relevant objects are: "toilet_bathtub_sink". 


## Example
User:
Request: animate the whale to flap its tail
Scene JSON: sceneName:LibraryTest,objects:[name:Compiler,name:Whale,children:[name:Armature,children:[name:Root,children:[name:Head,children:[name:Head_end],name:Spine1,children:[name:Spine2,children:[name:Spine3,children:[name:Spine4,children:[name:Tail,children:[name:Tail_end]]]],name:TopFlipper.L,children:[name:MidFlipper.L,children:[name:BottomFlipper.L,children:[name:BottomFlipper.L_end]]],name:TopFlipper.R,children:[name:MidFlipper.R,children:[name:BottomFlipper.R,children:[name:BottomFlipper.R_end]]]]]],name:Whale]]


Assistant:
Relevant objects: "Whale", "Tail", "Spine4", "Armature".

\end{lstlisting}
\lstset{style=mystyle}

\subsection{Skill Library}

\lstinputlisting[style=mystyle, caption={Skill Library's metaprompt},label=p3, frame=single]{Metaprompts/APIManager_CF.txt}
\lstset{style=mystyle}

\subsection{Builder}

\lstinputlisting[style=mystyle, caption={Builder's metaprompt},label=p4, frame=single]{Metaprompts/Builder_SA_SL.txt}
\lstset{style=mystyle}

\subsection{Inspector}

\lstinputlisting[style=mystyle, caption={Inspector's metaprompt},label=p5, frame=single]{Metaprompts/Inspector_SL.txt}
\lstset{style=mystyle}

\newpage
\section{User study Questionnaire and Interview Questions}
 Rate these statements (All 1 to 7 Likert scale from Strongly disagree to Strongly agree):
 
 \begin{itemize}
     \item I feel frustrated when I use the framework.
     \item I am satisfied with the outcome of this framework.
     \item I was able to iteratively improve the output of my generation.
     \item I was able to achieve what I had in mind with this framework. 
     \item The prompting process of creating 3D environments and objects enabled by the framework was intuitive.
     \item I will recommend this tool to a colleague.
     \item I will use this tool in my work.

 \end{itemize}

Semi-structured Interview questions:
\begin{itemize}
    \item What was your approach to prompting during this study?
    \item How would you compare this approach to 3D world creation to your typical experience of creating 3D environments? What was your prior experience? And how was this approach similar or different? When would you use this over your typical approach
    \item What were some surprising parts of this framework?
    \item What would you wish you could do that did not get to do? What didn't think you could do with this framework?
    \item What applications can you see this framework being used in?
\end{itemize}






\end{document}